\documentclass[11pt,a4paper]{article}
\pdfoutput=1
\usepackage{fancyhdr}  
\usepackage{jcappub}
\usepackage{bbm}
\usepackage{mathrsfs}
\usepackage{slashed}
\usepackage{caption}
\usepackage{epstopdf}
\usepackage[normalem]{ulem}
\usepackage[bottom]{footmisc}
\usepackage{subcaption}
\usepackage{bbold}
\usepackage{titlesec}
\usepackage{threeparttable}
\usepackage{booktabs}
\usepackage{changepage}
\usepackage[utf8]{inputenc}
\usepackage{dsfont}
\usepackage{grffile}
\usepackage{graphicx}  % needed for figures
\usepackage{dcolumn}   % needed for some tables
\usepackage{bm}        % for math
\usepackage{setspace}
\usepackage{amsmath,amssymb,setspace}
\usepackage{array}
\usepackage{booktabs}
\usepackage{indentfirst}
\usepackage{float}
\usepackage{lmodern}
\usepackage{multirow}
\usepackage{soul}
\usepackage[dvipsnames]{xcolor}
\usepackage{extpfeil}
\usepackage{enumitem} % customize \item
\usepackage{braket}
\usepackage{comment}
\usepackage{hyperref}
\makeatletter
\newcommand*\bigcdot{\mathpalette\bigcdot@{.5}}
\newcommand*\bigcdot@[2]{\mathbin{\vcenter{\hbox{\scalebox{#2}{$\m@th#1\bullet$}}}}}
\makeatother

\newcommand{\calO}{{\cal O}}
\newcommand{\calR}{{\mathcal{R}}}

\newcommand{\calP}{{\cal P}}
\newcommand{\be}{\begin{equation}}
\newcommand{\ee}{\end{equation}}

\newcommand{\ur}{\mathrm{r}}
\newcommand{\calW}{\mathcal{W}}

\usepackage{verbatim}
\usepackage{tikz}
\usetikzlibrary{arrows.meta}

\tikzset{%
  >={Latex[width=2mm,length=2mm]},
            base/.style = {rectangle, rounded corners, draw=black,
                           minimum width=4cm, minimum height=1cm,
                           text centered, font=\sffamily},
  activityStarts/.style = {base, fill=blue!30},
       startstop/.style = {base, fill=red!30},
    Blockmuth/.style = {base, fill=green!30},
         process/.style = {base, minimum width=2.5cm, fill=orange!15,
                           font=\ttfamily},
}

\usepackage{graphicx} 
\usepackage{epstopdf, epsfig}
\usepackage{amsmath}
\usepackage{comment}
\usepackage{color}
\usepackage{pifont}

\usepackage{xcolor}
\usepackage{mathrsfs,mathtools}
\usepackage{physics,amssymb}
\usepackage{siunitx}
\usepackage{bm}
\usepackage{soul}
\usepackage{cancel}
\usepackage{url}
\usepackage{longtable}
\usepackage{xspace}
\usepackage{acronym}
\usepackage{pifont}
\usepackage{xcolor}
\usepackage{tikz}

\hypersetup{colorlinks=true
,urlcolor=DARKBLUE
,anchorcolor=DARKBLUE
,citecolor=DARKBLUE
,filecolor=DARKBLUE
,linkcolor=DARKBLUE
,menucolor=DARKBLUE
,linktocpage=true
,pdfproducer=medialab
,pdfa=true
}

\definecolor{MONZA}{HTML}{CF000F}
\definecolor{DARKBLUE}{HTML}{00008b}
\definecolor{DARKMAGENTA}{HTML}{8b008b}

\newcommand{\fpk}{\mathrm{pk1}}
\newcommand{\spk}{\mathrm{pk2}}
\newcommand{\cut}{\mathrm{cut}}
\newcommand{\umid}{\mathrm{mid}}
\newcommand{\UV}{\mathrm{UV}}
\newcommand{\IR}{\mathrm{IR}}
\newcommand{\dip}{\mathrm{dip}}

\newcommand{\ue}{\mathrm{e}}

\newcommand{\us}{\mathrm{s}}

\newcommand{\beae}[1]{\begin{equation}\begin{aligned} #1 \end{aligned}\end{equation}}

\newcommand{\bae}[1]{\begin{align} #1 \end{align}}

\newcommand{\bde}[1]{\begin{dcases} #1 \end{dcases}}

\newcommand{\bme}[1]{\begin{multline} #1 \end{multline}}
\newcommand{\bmte}[1]{\begin{multlined}[t] #1 \end{multlined}}

\acrodef{CR}{constant-roll}
\acrodef{SR}{slow-roll}
\acrodef{CMB}{cosmic microwave background}
\acrodef{SIGW}{scalar-induced gravitational wave}
\acrodef{PBH}{primordial black hole}

\allowdisplaybreaks[4]

\title{\huge{Constant-Roll Inflation: Analytical Formulae for Power Spectrum and Implications for Induced Gravitational Waves}}

\author[a]{Hayato Motohashi,}
\author[b,c]{Shi Pi,}
\author[d,e]{Yuichiro Tada,}
\author[c]{and Jianing Wang}

\affiliation[a]{Department of Physics, Tokyo Metropolitan University,
1-1 Minami-Osawa, Hachioji, Tokyo 192-0397, Japan}
\affiliation[b]{Institute of Theoretical Physics, Chinese
Academy of Sciences, Beijing 100190, China}
\affiliation[c]{Kavli Institute for the Physics and Mathematics of the Universe (WPI), UTIAS, The University of Tokyo, Kashiwa, Chiba 277-8583, Japan}
\affiliation[d]{Department of Applied Physics, University of Fukui, \\
Bunkyo 3-9-1, Fukui 910-8507, Japan}
\affiliation[e]{Department of Physics, Rikkyo University, \\ 
3-34-1 Nishi-Ikebukuro, Toshima, Tokyo 171-8501, Japan}

\emailAdd{motohashi@tmu.ac.jp}
\emailAdd{shi.pi@itp.ac.cn} 
\emailAdd{ytada@u-fukui.ac.jp}
\emailAdd{jianing.wang@ipmu.jp}

\abstract{Constant-roll inflation provides a simple and analytically tractable framework for describing transient departures from slow roll, including non-attractor phases that can enhance the primordial curvature perturbation on small scales.
In this work, we investigate the curvature power spectrum generated in a slow-roll--constant-roll--slow-roll scenario, focusing on the positions and amplitudes of the two characteristic peaks associated with the two transitions. 
We show that, in the parameter range where both peaks are well separated and sufficiently pronounced, the underlying constant-roll parameters can be reconstructed from the peak positions and amplitudes without performing a brute-force parameter scan. 
In addition, we construct a smoothed analytic approximation to the power spectrum, designed for efficient estimates of scalar-induced gravitational waves and related phenomenological applications.
}

\begin{document}

\maketitle
\flushbottom

%%%%%%%%%%%%%%%%%%%%%%%%%%%%%%%%%%%%%%%%%%%%%%%%%%%%%%%%%%%%%%%%%%%%%%%%

\section{Introduction}

In the study of the early universe, one of the most successful ideas is the inflationary paradigm~\cite{Brout:1977ix,Starobinsky:1980te,Kazanas:1980tx,Guth:1980zm,Sato:1981qmu}, which posits a brief period of extremely rapid expansion before the radiation-dominated era.
The inflationary phase helps explain why the universe appears so homogeneous and isotropic on large scales, and it also provides a mechanism for generating the small-scale density fluctuations that later grow into galaxies and cosmic structures~\cite{Linde:1981mu,Ratra:1987rm,Linde:1983gd,Guth:1982ec,Kolb:1990vq,Freese:1990rb,Mukhanov:1990me,Kodama:1984ziu,Copeland:1997et}.
In the conventional \ac{SR} picture~\cite{Channuie:2014ysa,Senatore:2016aui,Brandenberger:2016uzh,Gron:2018rtj,vanHolten:2023uja,DiMarco:2024yzn,Casadio:2005em}, the inflaton field that drives inflation evolves slowly down its potential, and the rate of change of its velocity is negligible. This ensures a nearly constant Hubble parameter and leads to a nearly scale-invariant spectrum of primordial curvature perturbation~\cite{Mukhanov:1985rz,Sasaki:1986hm}, which is verified by recent observations ~\cite{COBE:1992syq,Bennett:1996ce,WMAP:2012fli,WMAP:2012nax,Planck:2018vyg,Planck:2019nip}. 

Although the standard \ac{SR} picture explains most cosmological observations remarkably well, it relies on simplifying assumptions that need not hold throughout the entire inflationary history.
In particular, during certain stages of inflation, such as when the inflaton passes over a flat or steep feature in its potential~\cite{Starobinsky:1992ts,Biagetti:2018pjj,Cole:2022xqc,Pi:2022zxs,Artymowski:2015ida,Bhaumik:2019tvl,Zhang:2021vak,Iacconi:2021ltm,Cai:2021zsp,Pi:2022ysn,Wang:2024nmd,Wang:2024wxq,Tomberg:2025fku,Escriva:2025ftp}, the \ac{SR} conditions can break down. 
\Ac{CR} inflation~\cite{Motohashi:2014ppa,Motohashi:2025qgd} provides a controlled framework for studying such departures from slow roll while remaining within the standard inflationary paradigm.
\Ac{CR} inflation naturally includes \ac{SR} as a limiting case but also encompasses more exotic situations like ultra-slow-roll inflation~\cite{Namjoo:2012aa,Martin:2012pe,Germani:2017bcs,Ballesteros:2017fsr,Hertzberg:2017dkh,Pattison:2018bct,Liu:2020oqe,Ballesteros:2020sre,Cheng:2021lif,Figueroa:2021zah,Mishra:2023lhe,Choudhury:2023hvf,Franciolini:2023agm,Karam:2022nym,Cheng:2023ikq,Firouzjahi:2023bkt,Ballesteros:2024zdp,Kawaguchi:2024rsv,Fujita:2025imc}, which can amplify the curvature perturbation on superhorizon scales. 

Several open issues remain in the study of \ac{CR} inflation. 
A comprehensive understanding of non-Gaussianities~\cite{Atal:2019cdz,Pi:2022ysn,Inui:2024sce,Ballesteros:2024pbe}, super-horizon evolution \cite{Leach:2000yw,Artigas:2024ajh,Cruces:2025typ}, higher-order correlators~\cite{Adshead:2013zfa,Passaglia:2018afq,Passaglia:2018ixg,Motohashi:2023syh}, and backreaction~\cite{Cheng:2021lif} are still being developed. 
\ac{CR} models embedded in modified gravity frameworks~\cite{Awad:2017ign,Motohashi:2017vdc,Odintsov:2017hbk,Mohammadi:2018zkf,Motohashi:2019tyj,Oliveros:2019xef,Oikonomou:2020tct,Oikonomou:2020oil,Shokri:2021rhy,Garnica:2021fuu,Shokri:2021jxh,Saleem:2023aof,ElBourakadi:2023ufz,Nojiri:2023nop,Kurt:2024isi,Huang:2024xqk,Keskin:2025bce} --- such as Palatini gravity~\cite{Antoniadis:2020dfq,Panda:2022can}, brane-world setups~\cite{Mohammadi:2020ftb,Stojanovic:2023dni,Baffou:2023uki,Djordjevic:2024fxy},
warm inflation~\cite{Oikonomou:2017bjx,Kamali:2019wdh,Mun:2021kzb,Saleem:2023aof,Biswas:2025vlz},  
or non-minimal couplings~\cite{Pi:2021dft,Liu:2024uiy} --- extend the scope of the scenario and may introduce distinctive observational signatures. 
Stochastic and other non-perturbative methods are increasingly recognized as important tools for analyzing \ac{CR} dynamics, especially in regimes where large fluctuations and superhorizon growth challenge the validity of standard perturbation theory \cite{Pattison:2021oen,Tomberg:2023kli,Artigas:2025nbm,Briaud:2025ayt}.

In this context, a transient departure from slow roll is essential~\cite{Motohashi:2017kbs}: the background can be modelled as an \ac{SR}--\ac{CR}--\ac{SR} sequence, in which a temporary \ac{CR} phase enhances small-scale curvature perturbation until the system returns to an \ac{SR} attractor. Phenomenologically,
generic predictions of inflationary scenarios capable of producing large scalar fluctuations include the emergence of a stochastic gravitational-wave background induced at second order by scalar perturbations, 
and a population of \acp{PBH} which are formed by the gravitational collapse around the overdensed peaks. 
Constraints or measurements of \ac{SIGW} spectrum and \ac{PBH} abundance can thus serve as probes of the primordial scalar spectrum and the underlying inflationary process. 
The dynamics in \ac{CR} models~\cite{Ito:2017bnn,Odintsov:2017qpp,Oikonomou:2017xik,Gao:2019sbz,MohseniSadjadi:2019vvs,Odintsov:2019ahz,Mohammadi:2019qeu,Guerrero:2020lng,Sadeghi:2021egp,Mohammadi:2022vru,Sasaki:2025zao} can play an important role in enhancing the curvature perturbation that could lead to \ac{PBH} formation and observable \ac{SIGW} signals~\cite{Saito:2008jc,Inomata:2017okj,Garcia-Bellido:2017aan,Di:2017ndc,Cai:2018dig,Bartolo:2018evs,Unal:2018yaa,Motohashi:2019rhu,Ragavendra:2020sop,Figueroa:2020jkf,Bhaumik:2020dor,Domenech:2020ers,Wang:2021kbh,Biagetti:2021eep,Kitajima:2021fpq,Geller:2022nkr,Cai:2022erk,Escriva:2022duf,Choudhury:2023hfm,Firouzjahi:2023lzg,Frosina:2023nxu,Bhattacharya:2023ysp,Tada:2023rgp,Qin:2023lgo,Domenech:2023dxx,Domenech:2024rks,Pi:2024lsu,Croney:2024xzw,Barker:2024mpz,Inui:2024fgk,Pi:2024ert,Firouzjahi:2025ihn,Domenech:2026nun}. 
Current and upcoming gravitational-wave detectors form a multi-band observational network. Ground-based detectors such as Advanced LIGO, Virgo, and KAGRA~\cite{LIGOScientific:2007fwp,Sutton:2008zza,LIGOScientific:2011hqo,Kawamura:2019jqt,LIGOScientific:2019lzm,LIGOScientific:2019obb,LIGOScientific:2019vic,LIGOScientific:2021yby,LIGOScientific:2024sgg}, together with future facilities such as Cosmic Explorer~\cite{Reitze:2019iox} and the Einstein Telescope~\cite{ET:2025xjr} are designed to probe the hundred-Hertz band.
Space-based missions such as LISA~\cite{LISAPathfinder:2017khw,LISAPathfinder:2019bgi,LISA:2022kgy,LISACosmologyWorkingGroup:2022kbp,LISACosmologyWorkingGroup:2023njw}, Taiji~\cite{TaijiScientific:2021qgx}, TianQin~\cite{TianQin:2015yph,TianQin:2020hid} can probe millihertz signals. 
Pulsar timing arrays, including NANOGrav~\cite{NANOGrav:2020spf,NANOGrav:2023gor}, EPTA~\cite{EPTA:2011kjn,Kramer:2013kea,EPTA:2015qep,EPTA:2023gyr}, IPTA~\cite{IPTA:2023ero,InternationalPulsarTimingArray:2023mzf}, PPTA~\cite{PPTA:2024xbe}, CPTA~\cite{Xu:2023wog} are sensitive to nanohertz gravitational waves. 
As this multi-band network prepares to yield a massive influx of data in the near future, efficiently scanning the parameter space for potential \ac{SIGW} signals will pose a formidable computational challenge, highlighting the need for analytic templates of the enhanced power spectrum.

Our primary aim is to provide an analytic understanding of the characteristic features of the curvature power spectrum generated in \ac{SR}--\ac{CR}--\ac{SR} models. 
Building on this understanding, we construct a smoothed expression for the power spectrum in which rapidly oscillating features are averaged out while the main peak structure is retained.
This expression allows the model parameters to be inferred from an enhanced power spectrum with two pronounced peaks, a situation that typically arises in non-attractor scenarios with the second slow-roll parameter $\eta$ usually taking values around $-6$ to $-4$. 
It also provides a useful tool for estimating the corresponding \ac{SIGW} signals.

The open-source code associated with this work is available from~\cite{CRsolver}. 
Both Mathematica and Python implementations allow users to infer the model parameters and plot the corresponding \ac{SIGW} spectrum. 
The required inputs are the frequency ratio and amplitude ratio of the two peaks, which are used to reconstruct the shape of the power spectrum. In addition, the position of the highest peak is also required. 
This is complemented by a parameter that determines the overall amplitude—either the highest peak amplitude of the power spectrum or its infrared (IR) normalization. 
Although our analytic reconstruction formulae are derived for the \ac{SR}--\ac{CR}--\ac{SR} model, the smoothed-spectrum construction developed here may also be useful as a phenomenological template for other mechanisms that generate scalar spectra with similar two-peak structures.

The paper is organized as follows. 
In Sec.~\ref{s:summary}, we list the main formulae developed in this work. Readers primarily interested in applications may skip the intermediate derivations. 
In Sec.~\ref{s:Features}, we present the general solution for the \ac{SR}--\ac{CR}--\ac{SR} model and analyze the power spectrum from the infrared (IR) to the ultraviolet (UV) regime. 
In particular, we derive useful estimates for the dip position, the peak positions, and their amplitudes. 
In Sec.~\ref{s:solvebN}, we explain how to infer the model parameters from the peak information. 
In Sec.~\ref{s:IGW}, we show how the smoothed power spectrum can be used to estimate the \ac{SIGW} spectrum and compare it with the result obtained from the exact spectrum. 
We conclude in Sec.~\ref{s:conclusion}.

\section{Executive Summary}\label{s:summary}

For the convenience of the readers, we summarize our main results below.
\begin{enumerate}
    \item We consider a transient \ac{CR} phase, during which the second slow-roll parameter $\eta=\dot{\epsilon}/(\epsilon H)$, defined in terms of the first slow-roll parameter $\epsilon=-\dot{H}/H^2$, takes the constant value $2\beta$. A dot denotes the derivative with respect to cosmic time throughout this paper.
    The \ac{CR} phase starts at conformal time $\tau_\us$ and ends at $\tau_\ue$, and is sandwiched between two standard \ac{SR} phases.

    \item Suppose that the positions and amplitudes of the two peaks of the curvature power spectrum, $\pqty{k_\fpk,A_\fpk=\calP_\calR(k_\fpk)}$ and $\pqty{k_\spk,A_\spk=\calP_\calR(k_\spk)}$, are specified. `$\fpk$' and `$\spk$' denote the first peak and second peak, respectively. These two peaks are associated with the transitions at $\tau_\us$ and $\tau_\ue$, respectively. 
    The \ac{CR} parameter $\beta$ is then obtained as a solution of
    \bae{
        \frac{A_\spk}{A_\fpk}=\pqty{\frac{k_\spk}{k_\fpk}\frac{x_\fpk(\beta)}{x_\spk(\beta)}}^{2\beta+6}\frac{f_2(\beta)}{f_1(\beta)},
    }
    where
    \beae{
        x_\fpk(\beta)&=1.69-0.362 \beta+0.0326 \beta^2, \\
        x_\spk(\beta) &= 2.14-0.145 \beta-0.406 \beta^2-0.235 \beta^3-0.0465 \beta^4,  \\
        f_1(\beta)&=1637.551 +2376.283 \beta +1310.923 \beta ^2+323.051 \beta ^3+ 29.949 \beta ^4,\\
        f_2(\beta)&=51.354 +89.605 \beta +60.791 \beta^2 +18.264 \beta^3 +2.086 \beta^4 .
    }
    The transition times $\tau_\us$ and $\tau_\ue$ are reconstructed by
    \bae{
        \tau_\us=-\frac{x_\fpk(\beta)}{k_\fpk} \qc
        \tau_\ue=-\frac{x_\spk(\beta)}{k_\spk}.
    }
    The method applies to $-3 \leq \beta \lesssim -2$ and a sufficiently long \ac{CR} duration, where both peaks are sufficiently high and well separated from each other. 
    The discussion for other $\beta$ values can be found in Sec.~\ref{s:Features}. The $e$-folding number of \ac{CR} stage is obtained as
    \bae{
    N_{\mathrm{CR}}=\ln\left(\frac{k_{\spk}}{k_{\fpk}}\frac{x_{\fpk}(\beta)}{x_{ \spk}(\beta)}\right) .
    }
    Instead, one can specify the IR amplitude $A_\IR$ and the ratios $A_\fpk/A_\spk$ and $k_\fpk/k_\spk$. 
    
    \item Define the three characteristic scales:
    \bae{
        k_{\cut,1}\coloneqq k_\fpk-\frac{\pi}{2\tau_\us}, \quad 
        k_{\cut,2}\coloneqq k_\spk-\frac
        {\pi}{2\tau_\ue}, \quad 
        k_\umid \coloneqq -\left(\frac{3-2\nu}{2\nu-1}\frac{d_0(\beta)}{d_2(\beta)}\right)^{1/2}\frac{1}{\tau_\ue},
    }
    where $\nu\coloneqq \abs{3/2+\beta}$ and the functions $d_i(\beta)$ are given in Eq.~\eqref{eq:around2ndpkcoeff}.
    The curvature power spectrum is then well approximated by the following piecewise formulae: if $k_\umid>k_{\cut,1}$ (see the left panel of Fig.~\ref{fig: approximate power}),
    \bae{\label{eq: power fitting 4}
        \calP_\calR(k)\approx\bde{
            \calP_\calR^\fpk(k), & k<k_{\cut,1}, \\
            A_\umid\pqty{\frac{k}{k_\umid}}^{3-2\nu}, & k_{\cut,1}\leq k<k_\umid, \\
            \calP_\calR^\spk(k), & k_\umid\leq k<k_{\cut,2}, \\
            A_\UV, & k_{\cut,2}\leq k,
        }
    }
    or if $k_\umid\leq k_{\cut,1}$ (see the right panel of Fig.~\ref{fig: approximate power}),
    \bae{\label{eq: power fitting 3}
        \calP_\calR(k)\approx\bde{
            \calP_\calR^\fpk(k), & k<k_{\cut,1}, \\
            \calP_\calR^\spk(k), & k_{\cut,1}\leq k<k_{\cut,2}, \\
            A_\UV, & k_{\cut,2}\leq k.
        }
    }
    Here, the first- and second-peak shapes $\calP_\calR^\fpk(k)$ and $\calP_\calR^\spk(k)$ are given by
    \bae{
        \calP_\calR^\fpk(k)\coloneqq A_\IR \qty(\frac{\tau_\ue}{\tau_\us})^{6+4 \beta} (-k\tau_\us)^{4 } \mathcal{W}_d^{-1}(\beta)\sum_{n=0}^7 \mathcal{W}_{2 n}(\beta)(-k\tau_\us)^{2 n}
    }
    and
    \bae{
        \calP_\calR^\spk(k)\coloneqq A_\UV     \sum_{n=0}^6 d_{2 n}(\beta) (-k\tau_\ue)^{2 n},
    }
    where the explicit forms of the functions $\calW_i(\beta)$ are given in Eq.~\eqref{eq:Yi}.
    The UV amplitude is calculated as
    \bae{
        A_\UV=A_\IR\pqty{\frac{\tau_\ue}{\tau_\us}}^{2\beta}.
    }
    The middle amplitude $A_\umid$ is given by
    \bae{
        A_\umid=A_\UV\bqty{d_0(\beta)+d_2(\beta)(-k_\umid\tau_\ue)^{2}}.
    }

    \item 
    Substituting the analytic formula, Eq.~\eqref{eq: power fitting 4} or Eq.~\eqref{eq: power fitting 3}, into Eqs.~\eqref{eq: OGWh2} and \eqref{eq:igw-general}, one can numerically evaluate the present-day density parameter of the \ac{SIGW}, as exemplified in Fig.~\ref{fig: OGW}.
    
\end{enumerate}

Steps 2--4 are implemented in our open-source code~\cite{CRsolver}.
In Mathematica, calculations of the \ac{SIGW} spectrum can be time-consuming. 
For the \ac{SIGW} spectrum generated from the exact \ac{CR} power spectrum, the computation using Wolfram Mathematica 14.2 takes about $10$ minutes, whereas using the smoothed power spectrum proposed in this work requires about $1$ minute on an Apple M4-based system with 16 GB memory running macOS Sequoia 15.3.1. 
The Python version of the \ac{SIGW} integration is based on the public code \textsc{SIGWfast}~\cite{Witkowski:2022mtg,SIGWfast}. 
A single calculation typically completes within $1$ second. 
After downloading the Python package, users can open the \texttt{.ipynb} file, specify the same input parameters below the global settings as in the Mathematica implementation, and select the power-spectrum type by setting the \texttt{Type} variable to either \texttt{exact} or \texttt{smoothed}.
The results are printed directly, together with plots of the power spectrum and the \ac{SIGW} spectrum. 
The functions specific to this work are included in the file \texttt{functionscr.py}.

\section{Power Spectrum and Detailed Features}\label{s:Features}

Constant-roll inflation has drawn attention because it provides a useful way to describe the inflaton dynamics near features in the potential or during transitions between distinct inflationary phases.
Instead of requiring the inflaton acceleration to be negligible, constant-roll inflation assumes a fixed relation between the acceleration and the velocity, $\ddot{\phi}/(H\dot{\phi})=\text{constant}\eqqcolon\tilde\beta$~\cite{Motohashi:2014ppa},
where $\phi$ is the inflaton field, $\dot{\phi}$ and $\ddot{\phi}$ are its first and second time derivatives, $H$ is the Hubble parameter, and $\tilde\beta$ is the original constant-roll parameter.
This opens the door to a richer set of possible behaviors; for example, the inflaton can temporarily speed up or slow down. 
In this section, we first present the general solution of the \ac{SR}--\ac{CR}--\ac{SR} model and then study the detailed features of the curvature power spectrum.

\subsection{General Solutions}\label{s:General Solutions}

\begin{figure}
\centering
\includegraphics[width=10cm]{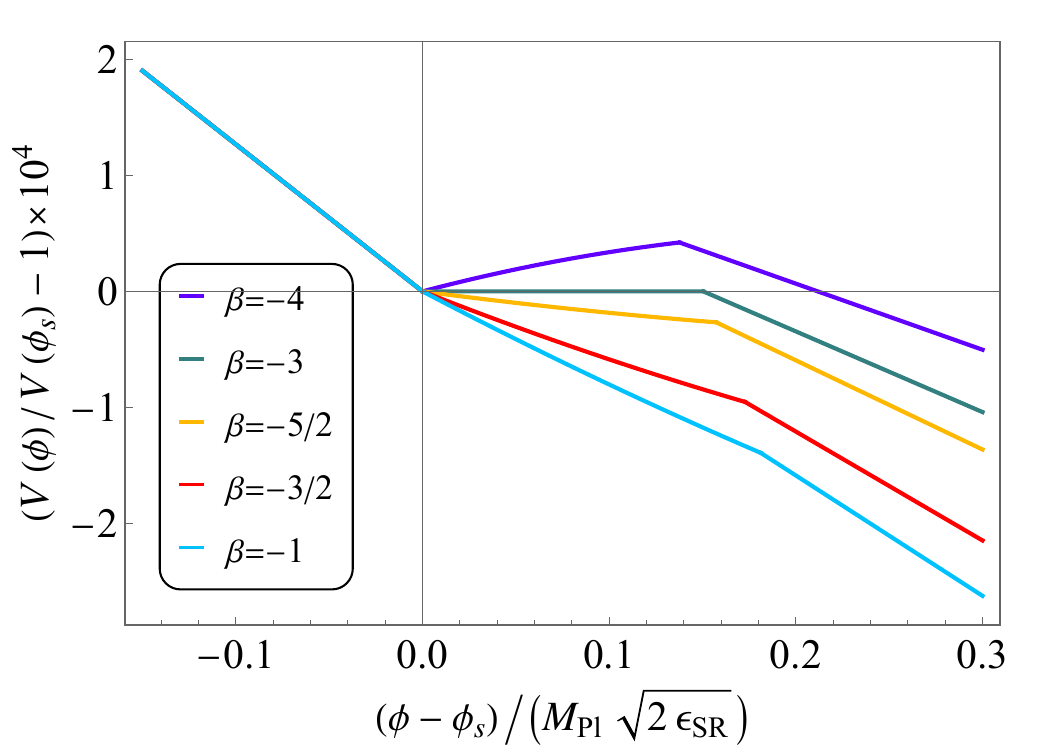}\\
\caption{Potentials of \ac{SR}--\ac{CR}--\ac{SR} inflation solved from Hamilton--Jacobi equation~\cite{Motohashi:2023syh}. We use $\epsilon_{\mathrm{SR}}=(H/M_{\mathrm{Pl}})^2 / (8\pi^2 A_\IR)$ where $H/M_{\mathrm{Pl}}\sim 10^{-5}$ 
and $A_\IR=2\times 10^{-9}$. 
}
\label{fig:potential}
\end{figure}

We begin with the definition of the model~\cite{Motohashi:2023syh}:
\begin{equation}
\begin{aligned}\label{def:CReta}
\eta\coloneqq \frac{\dot{\epsilon}}{\epsilon H}= \begin{cases}0, & \tau<\tau_\us, \\ 2 \beta, & \tau_\us \leq \tau<\tau_\ue, \\ 0, & \tau_\ue \leq \tau.\end{cases}
\end{aligned}
\end{equation}
$\epsilon=-\dot{H}/H^2$ denotes the first slow-roll parameter, and $\eta$ is the second slow-roll parameter. The conformal time $\tau$ is defined through $
\mathrm{d}\tau := \mathrm{d} t/a$, where $a$ is the scale factor.
For the ultra-slow-roll phase, $\eta=-6$ and hence $\beta=-3$. 
$\tau_\us$ denotes the transition from the first \ac{SR} phase to the \ac{CR} phase, while $\tau_\ue$ denotes the transition from the \ac{CR} phase to the second \ac{SR} phase, with $\tau_\us < \tau_\ue <0$.
As mentioned above, the constant-roll parameter is originally defined as $\ddot{\phi}/(H\dot{\phi})\eqqcolon\tilde{\beta}$~\cite{Motohashi:2014ppa}.
For a canonical scalar field, one has $\eta=2\ddot{\phi}/(H\dot{\phi})+2\epsilon$, and therefore the definition in Eq.~\eqref{def:CReta} implies $\beta=\tilde{\beta}+\epsilon$. 
Hence $\beta\simeq\tilde{\beta}$ when $\epsilon$ is negligible, or equivalently when the Hubble parameter is approximately constant, as assumed throughout this paper.
If one considers the time evolution of $H$, slightly different discussions can be found in Refs.~\cite{Motohashi:2014ppa,Ito:2017bnn,Anguelova:2017djf,Motohashi:2017aob,Gao:2017owg,Yi:2017mxs,Morse:2018kda,Motohashi:2019tyj,MohseniSadjadi:2019vvs,Lin:2019fcz,Gao:2019sbz,Nojiri:2023nop,Ahmadi:2023qcw,Inui:2024sce,Sadeghi:2024nvb,Wang:2024xdl,Motohashi:2025qgd,NooriGashti:2025enx}.
Solving Eq.~\eqref{def:CReta} under the assumption $\epsilon\ll1$, we have
\begin{equation}
\begin{aligned}
\epsilon
= \begin{cases}\epsilon_{\mathrm{SR}}, & \tau<\tau_\us, \\ \epsilon_{\mathrm{SR}}\left(\dfrac{\tau}{\tau_\us}\right)^{-2 \beta}, & \tau_\us \leq \tau<\tau_\ue, \\ \epsilon_{\mathrm{SR}}\left(\dfrac{\tau_\ue}{\tau_\us}\right)^{-2 \beta}. & \tau_\ue \leq \tau.\end{cases}
\end{aligned}
\label{eq:epsilon1}
\end{equation}
$\epsilon_{\mathrm{SR}}\ll 1$ is the first slow-roll parameter at the first \ac{SR} phase. The first slow-roll parameter $\epsilon$ decreases during the \ac{CR} phase, and small-scale modes would be highly enhanced. In Fig.~\ref{fig:potential}, we present the potential of the \ac{SR}--\ac{CR}--\ac{SR} model, where the $e$-folding number $N$ is defined through $\mathrm{d}N:=H\mathrm{d}t$. We set $N_\us=15$ and $N_\ue=15.2$ and start the calculation from $\phi(N=0)=0$ sufficiently before the first transition. The quantities $\phi_{\us,\ue}=\phi(N_{\us,\ue})$ denote the field values at the two transition points.

The  
canonical variable $v_k(\tau)$ 
is governed by the Mukhanov--Sasaki equation~\cite{Mukhanov:1985rz,Sasaki:1986hm}
\begin{equation}\label{eq:MSv}
v_k^{\prime \prime}+\left(k^2-\dfrac{z^{\prime \prime}}{z}\right) v_k=0, \quad \dfrac{z^{\prime \prime}}{z}= \begin{cases}\dfrac{2}{\tau^2}, & \text{for SR stages}, \\ \dfrac{\nu^2-1 / 4}{\tau^2},\quad \nu\coloneqq  |3 / 2+\beta| & \text{for CR stage}.\end{cases}
\end{equation}
A prime denotes the derivative with respect to conformal time.
The Mukhanov--Sasaki variable $v_k$ is related to the comoving curvature perturbation $\mathcal{R}_k$ by $v_k=z \mathcal{R}_k$ with $z=a \sqrt{2 \epsilon}$ (we adopt the reduced Planck units  
$M_{\mathrm{Pl}}=1$ afterwards).
The solution of Eq.~\eqref{eq:MSv} is 
\begin{equation}
\begin{aligned}
v_k(\tau)=\begin{cases}\sqrt{-k \tau} \left(C_1 H_{3/2}^{(1)}(-k \tau)+D_1 H_{3/2}^{(2)}(-k \tau)\right), & \text{first SR} ,\\ 
\sqrt{-k \tau} \left(C_2  H_{\nu}^{(1)}(-k \tau)+D_2 H_{\nu}^{(2)}(-k \tau)\right), & \text{CR} ,\\
\sqrt{-k \tau} \left(C_3 H_{3/2}^{(1)}(-k \tau)+D_3 H_{3/2}^{(2)}(-k \tau)\right). & \text{second SR}. \end{cases}
\end{aligned}\label{eq:vksol}
\end{equation}
Using the asymptotic behavior of the Hankel functions of order $\nu=3/2$,
\begin{equation}
H_{3 / 2}^{(1)}(x)\underset{~x\to \infty~}{\longrightarrow}-\sqrt{\frac{2}{\pi}}x^{-1 / 2}\left(1+\frac{i}{x}\right) e^{i x},\quad 
H_{3 / 2}^{(2)}(x)\underset{~x\to \infty~}{\longrightarrow}-\sqrt{\frac{2}{\pi}}x^{-1 / 2}\left(1-\frac{i}{x}\right) e^{-i x},
\end{equation}
one can match the first \ac{SR} phase with Bunch--Davies initial condition~\cite{Bunch:1978yq} to the infinite past
\begin{equation}
v_k \underset{~\tau\to -\infty~}{\longrightarrow} \frac{\mathrm{e}^{-i k \tau}}{\sqrt{2 k}}\left(1+\frac{i}{-k \tau}\right) ,
\end{equation}
which derives
\begin{equation}
C_1= -\frac{\sqrt{\pi}}{2\sqrt{ k}},\quad D_1=0.
\end{equation}
Define
\begin{equation}
\mathcal{A}_k\coloneqq \dfrac{C_3}{C_1},\quad \mathcal{B}_k\coloneqq  \dfrac{D_3}{C_1},
\end{equation}
and 
\begin{equation}
\kappa_\us \coloneqq  -k\tau_\us, \quad \kappa_\ue \coloneqq  -k\tau_\ue.
\end{equation}
The Israel junction conditions~\cite{Israel:1966rt}
\begin{equation}
\mathcal{R}_k\left(\tau_{\mathrm{s},\mathrm{e}}^{-}\right)=\mathcal{R}_k\left(\tau_{\mathrm{s},\mathrm{e}}^{+}\right), \quad \mathcal{R}_k^{\prime}\left(\tau_{\mathrm{s},\mathrm{e}}^{-}\right)=\mathcal{R}_k^{\prime}\left(\tau_{\mathrm{s},\mathrm{e}}^{+}\right)
\end{equation}
give
\begin{equation}
\begin{aligned}
    C_2&\bmte{= \frac{\pi e^{i \kappa_\us }}{2^{7/2} k^{1/2} \kappa_\us^{3/2}}\\
    \times\left\{ -2 \kappa_\us\left(1-i \kappa_\us\right) H_{\nu-1}^{(2)}(\kappa_\us)-\left( \left(3+2\beta-2\nu\right)\left(1-i \kappa_\us\right)-2 \kappa_\us^2\right) H_\nu^{(2)}(\kappa_\us)\right\},} \\
    D_2&\bmte{= \frac{-\pi e^{i \kappa_\us}}{2^{7/2} k^{1/2} \kappa_\us^{3/2}}\\
    \times\left\{ -2 \kappa_\us\left(1-i \kappa_\us\right) H_{\nu-1}^{(1)}(\kappa_\us)-\left( \left(3+2\beta-2\nu\right)\left(1-i \kappa_\us\right)-2 \kappa_\us^2\right) H_\nu^{(1)}(\kappa_\us)\right\} ,}
\end{aligned}
\label{eq:exact-C2D2}
\end{equation}
and
\begin{equation}
\begin{aligned}
    \mathcal{A}_k &\bmte{= \frac{e^{-i \kappa_\ue}}{2^{3/2} k^{-1/2} \kappa_\ue^{3/2}}\\
    \times\left\{ C_2 \left[ -2 \kappa_\ue\left(1+i \kappa_\ue\right) H_{\nu-1}^{(1)}(\kappa_\ue)+\left( -\left(3+2\beta-2\nu\right)\left(1+i \kappa_\ue\right)+2 \kappa_\ue^2\right) H_\nu^{(1)}(\kappa_\ue)\right]\right.\\
    +\left. D_2 \left[ -2 \kappa_\ue\left(1+i \kappa_\ue\right) H_{\nu-1}^{(2)}(\kappa_\ue)+\left( -\left(3+2\beta-2\nu\right)\left(1+i \kappa_\ue\right)+2 \kappa_\ue^2\right) H_\nu^{(2)}(\kappa_\ue)\right]\right\},} \\
    \mathcal{B}_k &\bmte{= \frac{e^{i \kappa_\ue}}{2^{3/2} k^{-1/2} \kappa_\ue^{3/2}}\\ 
    \times\left\{ C_2 \left[ -2 \kappa_\ue\left(1-i \kappa_\ue\right) H_{\nu-1}^{(1)}(\kappa_\ue)+\left( \left(3+2\beta-2\nu\right)\left(-1+i \kappa_\ue\right)+2 \kappa_\ue^2\right) H_\nu^{(1)}(\kappa_\ue)\right]\right.\\
    + 
    \left. D_2 \left[ -2 \kappa_\ue\left(1-i \kappa_\ue\right) H_{\nu-1}^{(2)}(\kappa_\ue)+\left( \left(3+2\beta-2\nu\right)\left(-1+i \kappa_\ue\right)+2 \kappa_\ue^2\right) H_\nu^{(2)}(\kappa_\ue)\right]\right\}.}
\end{aligned}
\label{eq:exact-CDAB}
\end{equation}
The final curvature power spectrum is
\begin{equation}
\mathcal{P}_\mathcal{R}(k)
\coloneqq  \frac{k^3}{2 \pi^2} |\mathcal{R}_k|^2 = A_\IR\left(\frac{\tau_\ue}{\tau_\us}\right)^{2\beta}\left| \mathcal{A}_k -\mathcal{B}_k \right|^2.
\label{eq:oriPowSpe}
\end{equation}
In deriving Eq.~\eqref{eq:oriPowSpe}, we used $a\approx -1/(H\tau)$ and assumed that the Hubble parameter $H$ is approximately constant during inflation. 
The overall normalisation $A_\IR$ corresponds to the IR limit, $k\ll-1/\tau_\us$, given by
\begin{equation}
A_\IR\coloneqq  \frac{(H/M_{\mathrm{Pl}})^2}{8 \pi^2 \epsilon_{\mathrm{SR}}} 
.
\end{equation}
We set $A_\IR=2\times 10^{-9}$ in figures unless otherwise specified. This choice is made to match the amplitude of the primordial curvature power spectrum inferred from cosmic microwave background (CMB) and large-scale structure observations~\cite{Planck:2018vyg,Planck:2018jri}.

Before proceeding to the next subsections, let us briefly review the crucial distinction between attractor and non-attractor solutions~\cite{Downes:2012xb,Karam:2017rpw,Morse:2018kda,Lin:2019fcz,Inui:2024sce}. 
An attractor solution is a dynamical trajectory for which the evolution loses memory of initial conditions and approaches a stable solution. 
A non-attractor solution, by contrast, retains sensitivity to initial conditions during inflation.
For our \ac{CR} model, $\eta=2\beta =2 \ddot{\phi}/(H \dot{\phi})+2 \epsilon$. 
Assuming $\epsilon\ll 1$, one finds $\ddot{\phi} /(H \dot{\phi})\approx\beta$, and hence $\dot{\phi} \propto a^\beta$. 
Therefore, $z\propto a^{1+\beta}$ for an approximately constant Hubble rate. 
The superhorizon solution of Eq.~\eqref{eq:MSv} is then $\mathcal{R}_k \sim A+B a^{-(2 \beta+3)}$. 
For $\beta>-3/2$, the second mode decays and $\mathcal{R}_k$ approaches a constant after horizon exit.
For $\beta<-3/2$, the second mode grows and $\mathcal{R}_k$ continues to evolve outside the horizon. 
Thus $\beta>-3/2$ corresponds to an attractor regime, while $\beta<-3/2$ corresponds to a non-attractor regime.

\begin{figure}
\centering
\includegraphics[width=12cm]{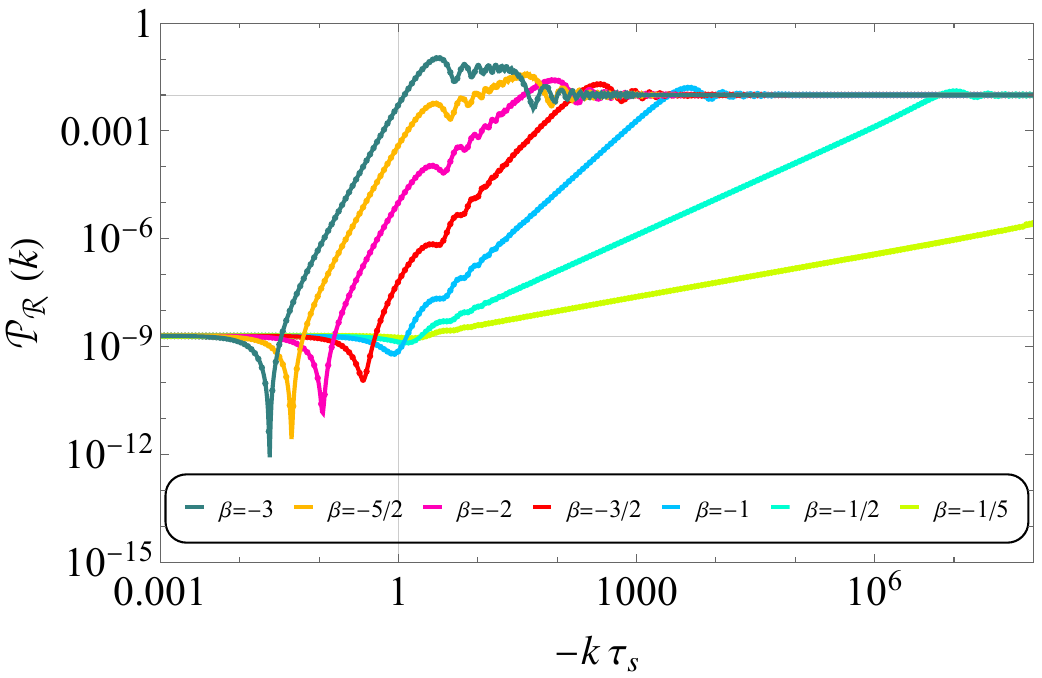}\\
\caption{
Dotted lines: numerical solutions calculated according to $z(\tau)$~\cite{Motohashi:2023syh}. Solid lines: analytical expression Eq.~\eqref{eq:oriPowSpe}. In this plot, we set $\left( \tau_\ue / \tau_\us\right)^{2 \beta}= 10^{-2}/ (2 \times  10^{-9})$.}
\label{fig:PRnumana}
\end{figure}

In the following subsections, we study the dip, peak, and enhancement slopes in the curvature power spectrum.
By ``dip'', we refer to the global minimum on the infrared (IR) side of the spectrum, located between the IR plateau and the enhancement associated with the first transition from \ac{SR} to \ac{CR}.
By ``peak'', we denote the first crest in each oscillation period associated with each transition.
Such features are generated by transient violations of slow roll and the resulting sharp changes in the background evolution~\cite{Motohashi:2017kbs}.
When the background changes non-adiabatically, curvature perturbations no longer simply freeze; instead, positive- and negative-frequency components mix and interfere, leaving oscillatory features in $\mathcal{P}_{\mathcal{R}}(k)$.

The Mukhanov--Sasaki equation can be regarded as a wave equation in Fourier space, where $z^{\prime \prime} / z$ plays the role of a time-dependent effective 
mass squared.
The rapid change of $z''/z$ across different phases induces a non-adiabatic transition for modes with the relevant values of $k$:
\begin{equation}
v_k^{(\mathrm{out})}=\alpha_k v_k^{(\mathrm{in})}+\beta_k v_k^{(\mathrm{in}) *},
\end{equation}
where $\alpha_k$ and $\beta_k$ are Bogoliubov coefficients ($|\alpha_k|^2-|\beta_k|^2=1$) describing mixing between the positive- and negative-frequency solutions.
Part of each mode is reflected by the change in the effective mass squared, while the remaining part is transmitted.
The interference between these two components modulates the final amplitude as 
\begin{equation}\label{eq:expforosl}
\frac{\left|v_k^{(\mathrm{out})}\right|^2}{\left|v_k^{(\mathrm{in})}\right|^2}\sim 1+2\left|\beta_k\right| \cos \theta_k,
\end{equation}
for $|\beta_k|\ll |\alpha_k|$, where $\theta_k=\arg(\alpha_k)-\arg(\beta_k)$ is the relative phase. 
The oscillatory term modulates the enhanced envelope and promotes some local crests and troughs to peaks and dips, as shown in Fig.~\ref{fig:PRnumana}.

\subsection{IR limit, Dip, and First Peak}

In this subsection, we derive approximate formulae for the IR regime in the non-attractor and attractor cases. 
Based on these IR approximations, we derive fitting formulae for the dip positions and the first peak position, both of which are generated by the transition from \ac{SR} to \ac{CR}.

\subsubsection[Non-attractor: $\beta<-3/2$]{\boldmath Non-attractor: $\beta<-3/2$}\label{ss:nonattIR}

\begin{figure}
\centering
\includegraphics[width=12cm]{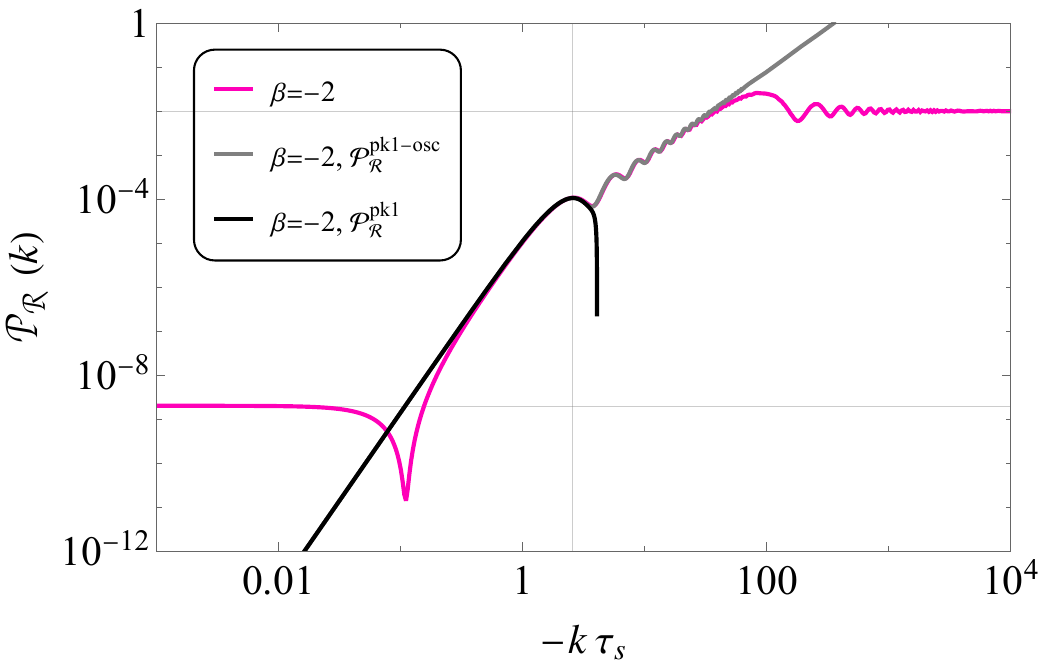}
\caption{The power spectra given by the exact solution Eq.~\eqref{eq:oriPowSpe}, the approximate formula Eq.~\eqref{eq: non-attractor approx}, and its Taylor expansion up to $\calO(\kappa_\us^{18})$ Eq.~\eqref{app:14}.
The approximation fails close to the attractor/non-attractor threshold $\beta=-3/2$. 
The vertical line shows the first-peak location $x_{\fpk}(\beta)$ given by Eq.~\eqref{eq:1stpkfit-nonatt}. 
}
\label{fig: non-attractor}
\end{figure}

For the non-attractor case, $\beta<-3/2$, the index of the Hankel function reads $\nu=-3/2-\beta$.
To obtain an approximate formula for the first peak around $\kappa_\us\sim1$, we replace the low-frequency Hankel functions by their asymptotic forms
\begin{equation}
    H_\nu^{(1)}(\kappa_\ue)\underset{\kappa_\ue \ll1}{\longrightarrow}-i\frac{\Gamma(\nu)}{\pi}\pqty{\frac{\kappa_\ue}{2}}^{-\nu} \qc
    H_\nu^{(2)}(\kappa_\ue)\underset{\kappa_\ue\ll1}{\longrightarrow}i\frac{\Gamma(\nu)}{\pi}\pqty{\frac{\kappa_\ue}{2}}^{-\nu}.
\label{eq:appHankel-o0}
\end{equation}
Keeping the leading terms in the limit $\tau_\ue/\tau_\us\ll1$, the power spectrum can be approximated as
\bme{\label{eq: non-attractor approx}
\calP_{\mathcal{R}}^{\fpk 
\text{-osc}}(k)\coloneqq  
A_\IR\left(\frac{\tau_\ue}{\tau_\us}\right)^{6+4\beta} \frac{2}{9}4^{-2-\beta}\beta^2\Gamma^2(\nu) \kappa_\us^{5+2\beta} \\
\times\left\{\kappa_\us^2J_\nu^2(\kappa_\us)+2 \kappa_\us J_\nu(\kappa_\us)J_{1+\nu}(\kappa_\us)+(1+\kappa_\us^2)J_{1+\nu}^2(\kappa_\us)\right\}.
}
We compare it with the exact solution Eq.~\eqref{eq:oriPowSpe} in Fig.~\ref{fig: non-attractor}. 
One may further expand Eq.~\eqref{eq: non-attractor approx} up to order $\kappa_\us^{18}$ to obtain a fitting formula for the first peak:
\begin{equation}
\mathcal{P}_{\mathcal{R}}^{\fpk}(k)\coloneqq  A_\IR\left(\frac{\tau_\ue}{\tau_\us}\right)^{6+4 \beta}  \kappa_\us^4 ~\mathcal{W}_d(\beta)^{-1}\sum_{n=0}^7 \mathcal{W}_{2 n}(\beta) \kappa_\us^{2 n},
\label{app:14}
\end{equation}
where the coefficients $\mathcal{W}_i(\beta)$ are given in Eq.~\eqref{eq:Yi}.
$\calP_{\mathcal{R}}^{\fpk 
\text{-osc}}(k)$ and $\mathcal{P}_{\mathcal{R}}^{\fpk}(k)$ are shown as the gray and black solid lines in Fig.~\ref{fig: non-attractor}, respectively.
These approximations fail for $\beta$ close to $-3/2$ and are accurate for $\beta\lesssim -1.8$. 
However, the prediction of the first-peak location from Eq.~\eqref{app:14} still works well, although the predicted amplitude can be much larger than the exact value. 
As $\beta$ approaches $-3/2$, the magnitude of the disagreement increases. 
To close this gap, one has to go beyond Eq.~\eqref{eq:appHankel-o0} and include subleading terms,
\begin{equation}
\begin{aligned}
    H_\nu^{(1)}(\kappa_\ue)&\underset{\kappa_\ue\ll1}{\longrightarrow} -i\frac{\Gamma(\nu)}{\pi}\pqty{\frac{\kappa_\ue}{2}}^{-\nu} -i\frac{\Gamma(-\nu)}{\pi}\pqty{\frac{\kappa_\ue}{2}}^{\nu}\cos(\pi \nu) +\frac{1}{\Gamma(1+\nu)}\pqty{\frac{\kappa_\ue}{2}}^{\nu}, \\
    H_\nu^{(2)}(\kappa_\ue)&\underset{\kappa_\ue\ll1}{\longrightarrow}  i\frac{\Gamma(\nu)}{\pi}\pqty{\frac{\kappa_\ue}{2}}^{-\nu}+i\frac{\Gamma(-\nu)}{\pi}\pqty{\frac{\kappa_\ue}{2}}^{\nu} \cos(\pi \nu)+\frac{1}{\Gamma(1+\nu)}\pqty{\frac{\kappa_\ue}{2}}^{\nu}.
    \end{aligned}
\label{eq:appHankel-o2}
\end{equation}
Considering the potentially intricate competition between the two modes around the critical case, together with the severe suppression of the first peak in realistic observations, we refrain from a detailed discussion here.

The first-peak position can be obtained by numerically solving for the maximum of the polynomial approximation in Eq.~\eqref{app:14}. For the non-attractor case $\beta<-3/2$, we solve for the maximum of
\begin{equation}\label{eq:solve1st-nonatt}
x^4 \sum_{n=0}^5 \mathcal{W}_{2 n}(\beta) x^{2 n}.
\end{equation}
Here we have neglected the last two terms in Eq.~\eqref{app:14}, because the higher-order terms are affected by subsequent oscillations; in particular, as $\beta\to -3/2$, some local oscillatory crests can even exceed the first crest associated with the SR--CR transition. The dominance of these higher-order terms indicates that the first peak becomes less sharply defined near the attractor/non-attractor boundary.
For $\beta<-3/2$, the numerical solution for the first-peak position can be fitted by the quadratic polynomial
\begin{equation}\label{eq:1stpkfit-nonatt}
-k_{\fpk}\tau_\us\approx 1.69-0.362 \beta+0.0326 \beta^2  
\eqqcolon x_{\fpk}(\beta),
\end{equation}
which is shown as the vertical line in Fig.~\ref{fig: non-attractor}.

\subsubsection[Attractor: $\beta>-3/2$]{\boldmath Attractor: $\beta>-3/2$}

\begin{figure}
\centering
\includegraphics[width=12cm]{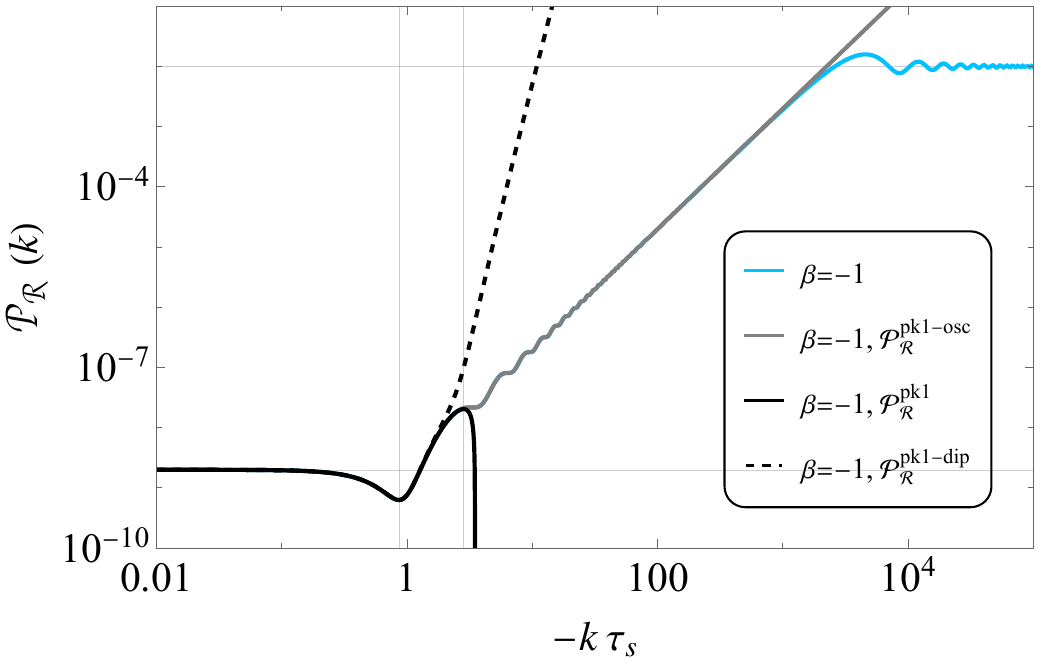}
\caption{Approximations for predicting the first-peak position and the dip position in the attractor case, with $\beta=-1$ shown as an example. 
$\mathcal{P}_{\mathcal{R}}^{\fpk\text{-osc}}(k)$, $\mathcal{P}_{\mathcal{R}}^{\fpk}(k)$, and $\mathcal{P}_{\mathcal{R}}^{\fpk\text{-dip}}(k)$ are given in Eqs.~\eqref{eq:eq:attractor approx}, \eqref{app:14attra}, and \eqref{app:8attra}, respectively. 
The vertical lines are given by Eq.~\eqref{eq:attr-kdip} for $-k_{\dip} \tau_\us$ and Eq.~\eqref{eq:1stpkfit-att} for $-k_{\fpk} \tau_\us$.}
\label{fig:attraIR}
\end{figure}

\begin{figure}
\centering
\includegraphics[width=10cm]{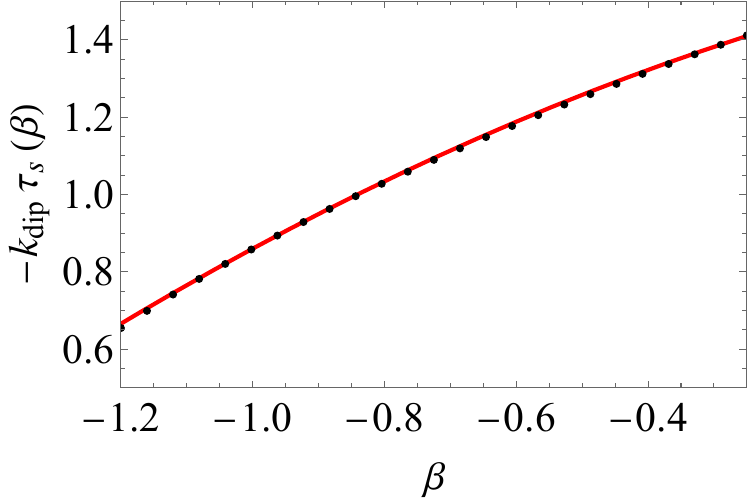}
\caption{Dip position for attractor case. Black dots: analytical expression Eq.~\eqref{eq:attr-kdip}. Red line: fitting formula Eq.~\eqref{eq:fittingattrdip}.}
\label{fig:attrdipfit}
\end{figure}

For the attractor case $\beta>-3/2$, the index of the Hankel function is $\nu=3/2+\beta$.
We repeat the same procedure as in the previous subsection, and obtain the gray line in Fig.~\ref{fig:attraIR}:
\bme{\label{eq:eq:attractor approx}
\calP_{\mathcal{R}}^{\fpk 
\text{-osc}}(k)\coloneqq 
A_\IR\frac{2}{9}\frac{4^{\beta +1} \Gamma^2 \left(\nu+1\right) }{(3+4 \beta(2+\beta))^2} \kappa_\us^{-3-2\beta}  \left(3+6 \beta-\left(\frac{\tau_\ue}{\tau_\us}\right)^2 \kappa_\us^2(1+\beta)\right)^2\\
\times\left\{\kappa_\us^2 \left(1+\kappa_\us^2\right) J^2_{\nu+1}(\kappa_\us)-2 \kappa_\us \left(3+2 \beta +2 (\beta +1) \kappa_\us^2\right) J_{\nu}(\kappa_\us) J_{\nu+1}(\kappa_\us)\right. \\
+
\left. \left((2 \beta +3)^2+(4 \beta  (\beta +2)+3) \kappa_\us^2 +\kappa_\us^4\right) J^2_{\nu}(\kappa_\us) \right\}.
}
Again, we can expand Eq.~\eqref{eq:eq:attractor approx} as polynomial:
\begin{equation}
\mathcal{P}_{\mathcal{R}}^{\fpk}(k)\coloneqq  A_\IR\left[ 1+\mathcal{Y}_0(\beta)^{-1}\sum_{n=1}^7 \mathcal{Y}_{2 n}(\beta) \kappa_\us^{2 n}\right],
\label{app:14attra}
\end{equation}
where the coefficients $\mathcal{Y}_i$ are listed in Eq.~\eqref{eq:appattr-18}. 
Equations~\eqref{eq:eq:attractor approx} and \eqref{app:14attra} are accurate for $\beta \gtrsim -1.25$.
Let us define $\mathcal{P}_{\mathcal{R}}^{\fpk\text{-}\dip}(k)$ using the first five leading terms of Eq.~\eqref{app:14attra}:
\begin{equation}
\mathcal{P}_{\mathcal{R}}^{\fpk\text{-}\dip}(k)\coloneqq  A_\IR\left[ 1+\mathcal{Y}_0(\beta)^{-1}\sum_{n=1}^4 \mathcal{Y}_{2 n}(\beta) \kappa_\us^{2 n}\right],
\label{app:8attra}
\end{equation}
We can estimate the dip position as
\begin{equation}
0=\eval{\dv{\calP_\calR^{\fpk\text{-dip}}}{k}}_{k=k_\dip}\propto\sum_{n=1}^4 n \cdot \mathcal{Y}_{2 n}(\beta) \left(-k_{\dip} \tau_\us\right)^{2 (n-1)},
\end{equation}
which is a cubic equation of $\left(-k_{\operatorname{dip}} \tau_\us\right)^2$.
The solution is given by
\begin{equation}\label{eq:attr-kdip}
-k_{\operatorname{dip}} \tau_\us =\left(-\frac{\mathcal{Y}_6}{4 \mathcal{Y}_8} -\frac{s_2}{6\cdot 2^{2/3} \mathcal{Y}_8}s^{-1/3} +\frac{1}{12 \cdot 2^{1/3} \mathcal{Y}_8}s^{1/3} \right)^{1/2} ,
\end{equation}
where
\begin{equation}
\begin{aligned}
s(\beta)&= \sqrt{s_3^2+4 s_2^3}+s_3,\\
s_3(\beta)&= 54\left(-8 \mathcal{Y}_2 \mathcal{Y}_8^2+ 4 \mathcal{Y}_4 \mathcal{Y}_6 \mathcal{Y}_8- \mathcal{Y}_6^3 \right),\\
s_2(\beta)&= 24 \mathcal{Y}_4 \mathcal{Y}_8-9 \mathcal{Y}_6^2.
\end{aligned}
\end{equation}
The prediction of Eq.~\eqref{eq:attr-kdip} is shown as the vertical line in Fig.~\ref{fig:attraIR}. 
We provide a fitting formula for Eq.~\eqref{eq:attr-kdip}, especially for $-1.2<\beta<-0.2$:
\begin{equation}\label{eq:fittingattrdip}
-k_{\dip} \tau_\us \simeq1.53+0.42 \beta-0.25 \beta^2,
\end{equation}
whose trend is shown in Fig.~\ref{fig:attrdipfit}. 

The first-peak position in the attractor case can be obtained in the same way as in the non-attractor case. 
Using Eq.~\eqref{app:14attra}, we solve for the maximum of
\begin{equation}\label{eq:solve1st-att}
\sum_{n=1}^7 \mathcal{Y}_{2 n}(\beta)  x^{2 n}.
\end{equation}
For $\beta>-3/2$, the numerical solution for the first-peak position is fitted by
\begin{equation}\label{eq:1stpkfit-att}
-k_{\fpk}\tau_\us\approx 3.52+0.496 \beta-0.194 \beta^2,
\end{equation}
which is shown as the vertical line in Fig.~\ref{fig:attraIR}.

\begin{figure}
\centering
\includegraphics[width=10cm]{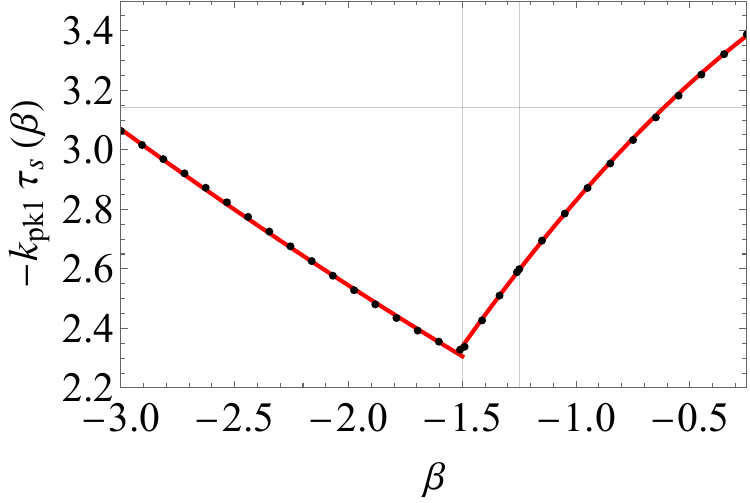}
\caption{First-peak position as a function of $\beta$. Black dots: numerical results from Eqs.~\eqref{eq:solve1st-nonatt} and \eqref{eq:solve1st-att}. Red lines: fitting formulae in Eqs.~\eqref{eq:1stpkfit-nonatt} and \eqref{eq:1stpkfit-att}.}
\label{fig:fit1stpk}
\end{figure}

Although Eqs.~\eqref{app:14} and \eqref{app:14attra} become inaccurate approximations to the first-peak amplitude when $\beta$ approaches $-3/2$, they still predict the first-peak location well. Therefore, the fitting formulae in Eqs.~\eqref{eq:1stpkfit-nonatt} and \eqref{eq:1stpkfit-att} can still be used around $\beta=-3/2$, as shown in Fig.~\ref{fig:fit1stpk}. 

\subsection{Enhancement}

In this section, we review the $k^4$ enhancement generated by the first transition and the $k^{3-2\nu}$ enhancement associated with the \ac{CR} stage. 
Systematic explanations are provided in Refs.~\cite{Motohashi:2014ppa,Byrnes:2018txb,Ozsoy:2019lyy,Artigas:2024ajh}. 
The presence of the $k^4$ enhancement enables a reliable prediction of the dip position over a wide range of $\beta$ values.
The $k^{3-2 \nu}$ enhancement plays a crucial role in constructing the smoothed power spectrum in Sec.~\ref{s:IGW}.
To begin this subsection, we introduce the 
$e$-folding number of the \ac{CR} phase for convenience:
\begin{equation}\label{def:eN}
\frac{\tau_\ue}{\tau_\us}=e^{-N_{\mathrm{CR}}}, \quad N_{\mathrm{CR}}>0.
\end{equation}
The amplitude of the curvature power spectrum corresponding to the second \ac{SR} phase is
\begin{equation}
A_\UV\coloneqq  A_\IR\left(e^{-N_{\mathrm{CR}}}\right)^{2 \beta}.
\label{def:AUV}
\end{equation}

\subsubsection[$k^4$-enhancement]{\boldmath $k^4$-enhancement}

\begin{figure}
\centering
\includegraphics[width=12cm]{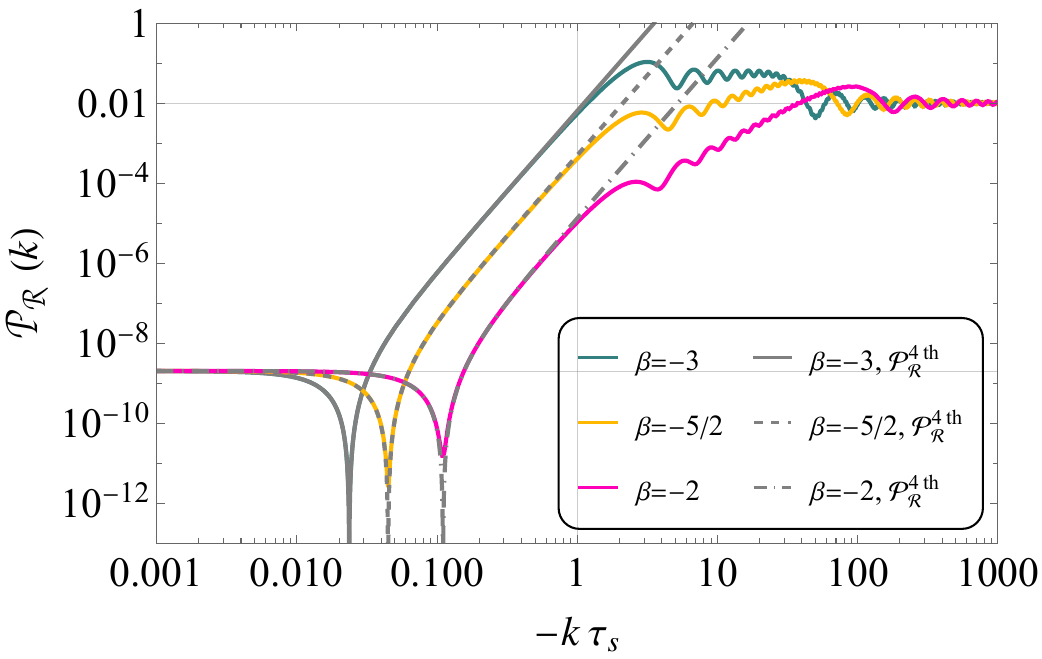}
\caption{Comparison of the full solution with the approximation in Eq.~\eqref{eq:fdip}. For the gray dashed line $\beta=-5/2$, the factor $1-\nu^2$ is cancelled to regularize the infinitesimal quantity in the denominator of Eq.~\eqref{eq:fdip}.}
\label{fig:k4thline}
\end{figure}

\begin{figure}
\centering
\includegraphics[width=11cm]{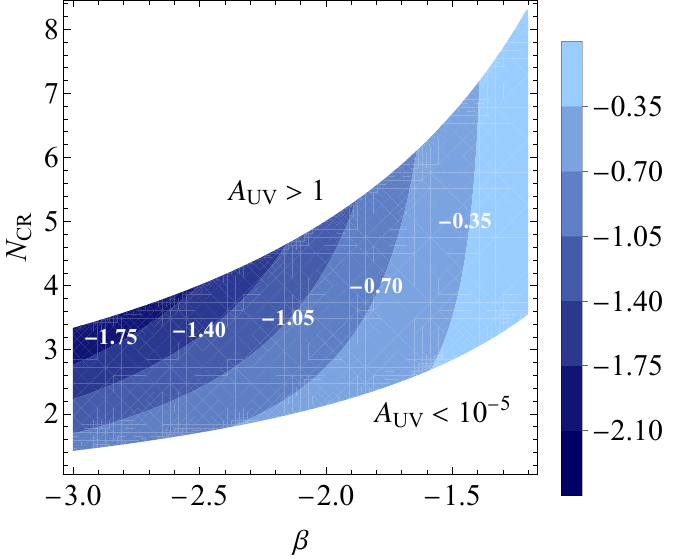}
\caption{Contour plot of $\log_{10}\left(-k_\dip \tau_\us\right)$, Eq.~\eqref{eq:dip}, valid for $-3\leq\beta\lesssim-1.2$. 
The plot is restricted to $10^{-5} \leq A_\UV \leq 1$, the range relevant for \ac{PBH} production or \ac{SIGW} observations.
The deviation of Eq.~\eqref{eq:dip} from the exact result is always within $10\%$.}
\label{fig:dipcontour}
\end{figure}

Mode mixing at the first transition produces a superhorizon transfer function whose small-$k$ behavior gives $\calP_{\mathcal{R}}(k) \propto k^4$~\cite{Leach:2001zf}. 
To analyze the long-wavelength modes, we expand $\mathcal{A}_k-\mathcal{B}_k$ in Eq.~\eqref{eq:exact-CDAB} around $\kappa_\us \to 0$. Keeping the first two leading orders gives
\begin{equation}
\mathcal{A}_k-\mathcal{B}_k \to \frac{e^{-(2(3+\nu)+1/2)N_{\mathrm{CR}}}}{2^5 ~3\nu (1-\nu^2)} \left[p_0\left(\beta, N_{\mathrm{CR}}\right)+p_2\left(\beta, N_{\mathrm{CR}}\right) \kappa_\us^2\right],
\label{eq:A-B}
\end{equation}
where
\begin{equation}
\begin{aligned}
p_0\left(\beta, N_{\mathrm{CR}}\right)&=\bmte{-4 (1-\nu^2) (2 \beta +2 \nu +3) (2 \beta -2 \nu -3) e^{(3 \nu +5) N_{\mathrm{CR}}}\\ 
+4 (1-\nu^2)  (2 \beta +2 \nu -3) (2 \beta -2 \nu +3) e^{(\nu +5) N_{\mathrm{CR}}},} \\
p_2\left(\beta, N_{\mathrm{CR}}\right)&=\bmte{ (\nu +1) (2 \beta +2 \nu -3) (2 \beta +2 \nu +3) e^{3 (\nu +1) N_{\mathrm{CR}}}\\
-2 (1-\nu^2) (2 \beta -2 \nu-3) (2 \beta +2 \nu +3) e^{(3 \nu +5) N_{\mathrm{CR}}}\\
+2 (1-\nu^2) (2 \beta -2 \nu +3) (2 \beta +2 \nu -3) e^{(\nu +5) N_{\mathrm{CR}}}\\
- (\nu -1) (2 \beta -2 \nu -3) (2 \beta +10 \nu +15) e^{(3 \nu +5) N_{\mathrm{CR}}}\\
- (\nu +1) (2 \beta +2 \nu -3) (2 \beta -10 \nu +15) e^{(\nu +5) N_{\mathrm{CR}}}\\
+ (\nu -1) (2 \beta -2 
\nu -3) (2 \beta -2 \nu +3) e^{(\nu +3) N_{\mathrm{CR}}}.}
\end{aligned}
\end{equation}
This can be used to estimate the dip position.
Considering Eq.~\eqref{eq:oriPowSpe}, the $k^4$ enhancement is approximately given by
\begin{equation}
\mathcal{P}_{\mathcal{R}}^{4\mathrm{th}}(k)\coloneqq 
A_\IR~ \frac{e^{-(4 (\nu  +3)+1+2\beta) N_{\mathrm{CR}} } }{ 2^{10}~3^2 \left(\nu   \left(1-\nu  ^2\right)\right)^2} \left[p_0\left(\beta, N_{\mathrm{CR}}\right)+p_2\left(\beta, N_{\mathrm{CR}}\right) \kappa_\us^2 \right]^2,
\label{eq:fdip}
\end{equation}
Figure~\ref{fig:k4thline} compares this approximation with the full solution. 
The higher-order terms in Eq.~\eqref{eq:A-B} affect only the dip amplitude, while the dip position can be obtained from Eq.~\eqref{eq:fdip} by requiring the bracketed factor to vanish:
\begin{equation}
-k_{\dip} \tau_\us=\left(-\frac{p_0(\beta,N_{\mathrm{CR}})}{p_2(\beta,N_{\mathrm{CR}})}\right)^{1/2}.
\label{eq:dip}
\end{equation}
This formula is not restricted to the non-attractor case, but applies whenever the $k^4$ enhancement is significant.
In the attractor case, Eq.~\eqref{eq:attr-kdip} shows that $-k_{\operatorname{dip}} \tau_\us$ depends mainly on $\beta$, whereas in the non-attractor case, Eq.~\eqref{eq:dip} shows that it also depends on $N_{\mathrm{CR}}$.
This behavior is illustrated in Fig.~\ref{fig:dipcontour}. 
The contour lines twist from a horizontal orientation, indicating a strong dependence on $N_{\mathrm{CR}}$, to a vertical orientation, indicating a weak dependence on $N_{\mathrm{CR}}$, as $\beta$ increases.

\subsubsection[$k^{3-2 \nu}$-enhancement]{\boldmath $k^{3-2 \nu}$-enhancement}

We now review the slope associated with the pure \ac{CR} stage. In Eq.~\eqref{eq:vksol}, $H_\nu^{(1)}$ corresponds to the positive-frequency mode in the Bunch--Davies vacuum (incoming modes in the distant past).
$H_\nu^{(2)}$ corresponds to the negative-frequency component (the ``reflected'' or excited mode) generated by mode mixing. 
If the Bunch--Davies vacuum is matched directly to the \ac{CR} stage without mode mixing, one has $v_k(\tau)\propto H_\nu^{(1)}(-k\tau)$. 
Using $\lim_{x\to0}H_\nu^{(1)}(x)\propto x^{-\nu}$, one obtains $\mathcal{P}_{\mathcal{R}}(k)\propto k^{3-2\nu}$~\cite{Motohashi:2014ppa}. 
For $3-2\nu>0$ ($-3<\beta<0$), this gives an enhancement due to the \ac{CR} phase, and the peak associated with the second transition can become higher than the one generated by the first transition. 

\subsection{UV limit and Second Peak}

\begin{figure}
\centering
\includegraphics[width=7.8cm]{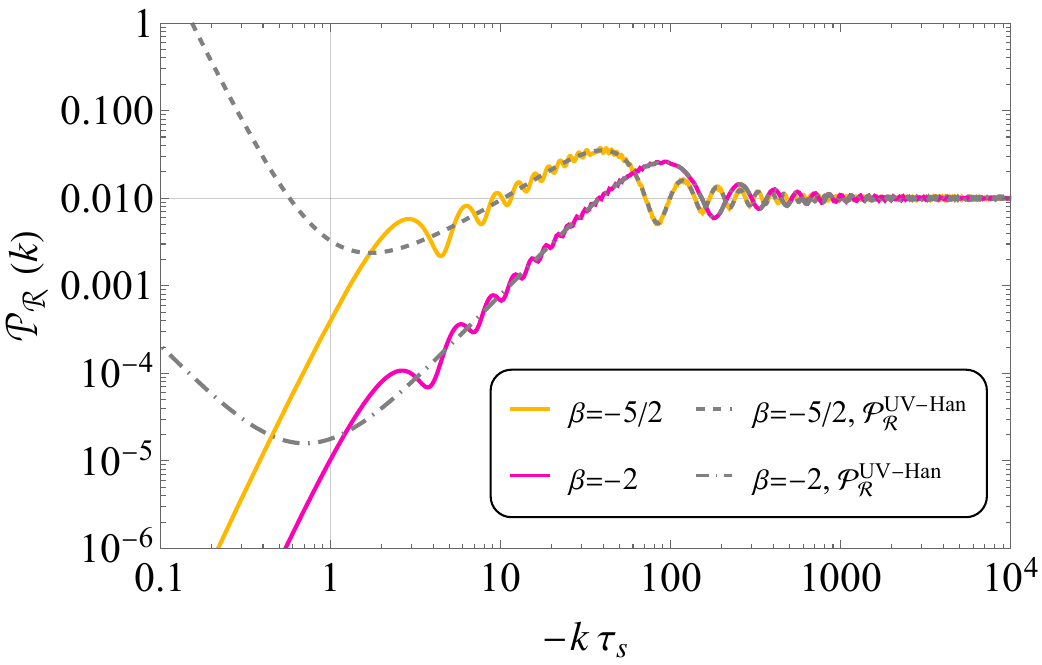}\includegraphics[width=7.8cm]{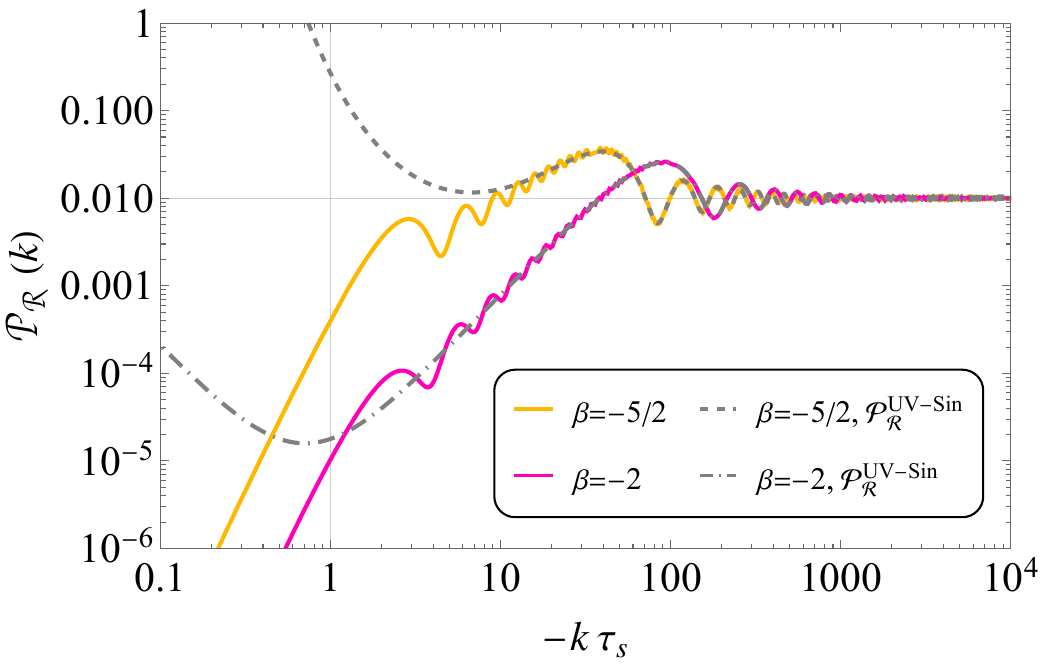}
\caption{Comparison of the UV approximations (dashed and dot-dashed) Eq.~\eqref{eq:UV-app-1} (left) or Eq.~\eqref{eq:2ndpeaktrig} (right) with the full expressions (solid lines) for $\beta=-5/2$ and $-2$. 
}
\label{fig:UVenv}
\end{figure}

As can be seen from Eqs.~\eqref{eq:exact-C2D2} and \eqref{eq:exact-CDAB}, the mode function contains Hankel functions with two types of arguments: the high-frequency variable $\kappa_\us$ and the low-frequency variable $\kappa_\ue$.
To estimate the second peak in the power spectrum, we neglect the high-frequency oscillations. 
More explicitly, the asymptotic phases of $H_\nu^{(1)}(z)$ and $H_\nu^{(2)}(z)$ behave as $\exp(i z)$ and $\exp(-i z)$, respectively. 
We therefore expand Eq.~\eqref{eq:exact-CDAB} by keeping track only of the phases $\exp(\pm i\kappa_{\us,\ue})$, and identify the lowest-frequency oscillatory contributions, while ignoring the slower amplitude modulation. 
In this decomposition, $\mathcal{A}_k$ contains a non-oscillating contribution, whereas the lowest-frequency oscillatory contribution in $\mathcal{B}_k$ is proportional to $\exp(2i\kappa_\ue)$. 
Keeping only the corresponding terms in $\mathcal{A}_k$ and $\mathcal{B}_k$ in Eq.~\eqref{eq:oriPowSpe}, we obtain
\bme{
\mathcal{P}_{\mathcal{R}}^{\text{UV-Han}}(k) \\ 
\coloneqq\frac{\pi^2 A_\UV}{2^6 \kappa_\us \kappa_\ue} \left|\left( Q_1 H_{\beta+\frac{1}{2}}^{(1)}\left(\kappa_\ue\right) +Q_2 H_{\beta+\frac{3}{2}}^{(1)}\left(\kappa_\ue\right)  \right)\left( Q_3 H_{\beta+\frac{1}{2}}^{(1)}\left(\kappa_\us\right) +Q_4 H_{\beta+\frac{3}{2}}^{(1)}\left(\kappa_\us\right)  \right)\right|^2 ,
\label{eq:UV-app-1}
}
where
\begin{equation}
Q_1= -2 \kappa_\ue \cos(\kappa_\ue)+2 \sin(\kappa_\ue),\quad Q_2= -2 \kappa_\ue \sin(\kappa_\ue),\quad  Q_3= i - \kappa_\us,\quad   Q_4= -i \kappa_\us.
\end{equation}
Eq.~\eqref{eq:UV-app-1} gives the envelope associated with the $\kappa_\ue$ oscillation.
Approximating the Hankel functions in Eq.~\eqref{eq:UV-app-1} for large argument, the UV limit can be further organized into a sinusoidal function:
\begin{equation}\label{eq:2ndpeaktrig}
\mathcal{P}_{\mathcal{R}}^{\text{UV-Sin}}(k)\coloneqq  A_\UV ~ \mathcal{S}_\us ~  \mathcal{S}_\ue, \quad \mathcal{S}_\ue =\mathcal{A}\sin\left( -2 \kappa_\ue + \Phi\right)+\theta. 
\end{equation}
We perform a third-order expansion of the Hankel functions in Eq.~\eqref{eq:UV-app-1} in the regime of large arguments to obtain Eq.~\eqref{eq:2ndpeaktrig}:
\begin{equation}
\begin{aligned}
& H_\nu^{(1)}\left(x\right) \underset{~x\to \infty~}{\longrightarrow} \frac{9+16 i x-128 x^2-8(5+8 i x) \nu^2+16 \nu^4}{64 \sqrt{2 \pi} x^{5 / 2}} e^{i(x+\pi(3-2 \nu)/4)},\\
& H_\nu^{(2)}\left(x\right) \underset{~x\to \infty~}{
\longrightarrow}\frac{-9+16 i x+128 x^2+8(5-8 i x) \nu^2-16 \nu^4}{128\sqrt{\pi} x^{5 / 2}}e^{i(-x+\pi \nu/2)}\left(1+i\right) .
\end{aligned}
\end{equation}
The goal of this step is to express the Hankel functions as power-law series, preferably with integer powers, so that explicit formulae for the second-peak location can be obtained later. 
The factor $\mathcal{S}_\us$ encodes the modulation associated with the high-frequency variable $\kappa_\us$, while $\mathcal{S}_\ue$ contains the lower-frequency oscillation associated with $\kappa_\ue$. 
The coefficients $\mathcal{A}$, $\Phi$, and $\theta$ can all be expanded as series in $\kappa_\ue$; their explicit expressions are given in Eq.~\eqref{eq:2ndpeaktrigcoeff}.
In Fig.~\ref{fig:UVenv}, we compare Eq.~\eqref{eq:UV-app-1} with Eq.~\eqref{eq:2ndpeaktrig}. The two approximations agree well around the second peak.

\begin{figure}
\centering
\includegraphics[width=12cm]{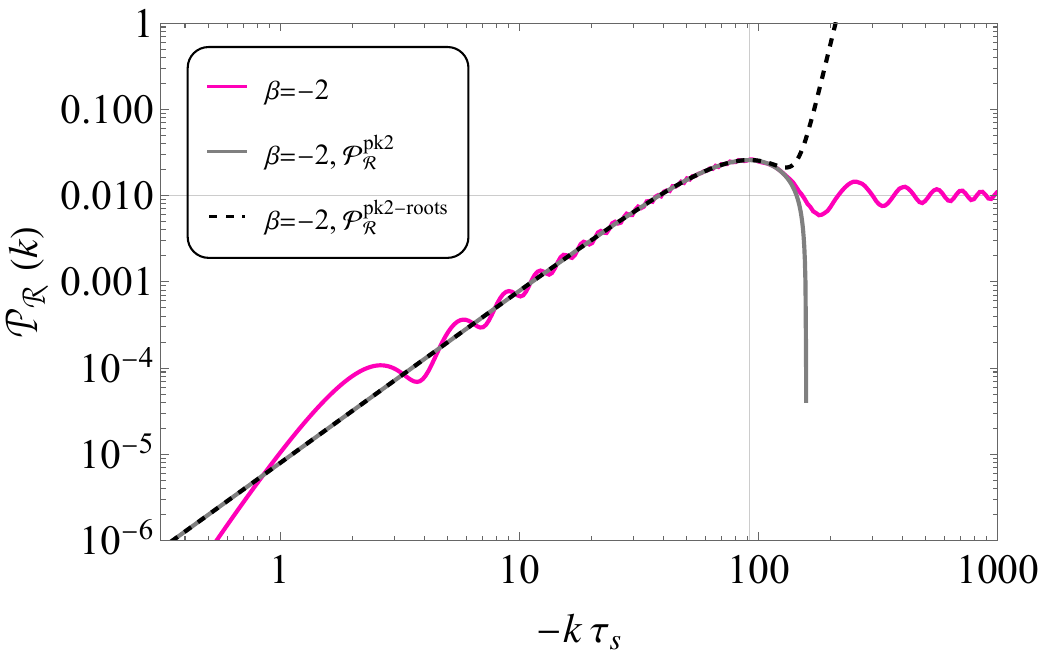}
\caption{Approximations around the second peak. The gray solid line shows Eq.~\eqref{eq:2ndpeakpl}, the black dashed line shows the truncated expression $\mathcal{P}_{\mathcal{R}}^{\spk\text{-roots}}(k)$, and the vertical line indicates the estimated second-peak position.
The horizontal line is $A_\UV=10^{-2}$. 
Eq.~\eqref{eq:2ndpeakpl} becomes negative on scales much larger than the second peak.
}
\label{fig:solve-2nd-n2}
\end{figure}

To determine the position of the second peak, we expand the sinusoidal function in Eq.~\eqref{eq:2ndpeaktrig} in powers of $\kappa_\ue$ around the second-peak region. 
Different from the large-argument (UV) expansion performed in Eqs.~\eqref{eq:UV-app-1} and \eqref{eq:2ndpeaktrig}, the current expansion is instead carried out in the small-argument (IR) regime. This is justified because the second peak lies closer to the IR side of the spectrum in Eq.~\eqref{eq:2ndpeaktrig}.
By retaining the results of the first twelve orders (gray solid line in Fig.~\ref{fig:solve-2nd-n2}), we find a good approximation around the second peak: 
\begin{equation}\label{eq:2ndpeakpl}
\mathcal{P}_{\mathcal{R}}^{\spk}(k)\approx
A_\UV\sum_{n=0}^{6} d_{2n}(\beta)\kappa_\ue^{2n}.
\end{equation}
The coefficients $d_{2n}(\beta)$ are listed in Eq.~\eqref{eq:around2ndpkcoeff}.
However, finding the extrema of Eq.~\eqref{eq:2ndpeakpl} requires solving the fifth-order equation
\begin{equation}\label{eq:5throot}
d_2 +2 d_4 x+3 d_6 x^2+4 d_8 x^3+5 d_{10} x^{4}+6 d_{12}x^{5}=0,
\end{equation}
where $x\coloneqq  (-k_{\spk} \tau_\ue)^2$. 
This equation has no simple closed-form solution in radicals.

\begin{figure}
\centering
\includegraphics[width=10cm]{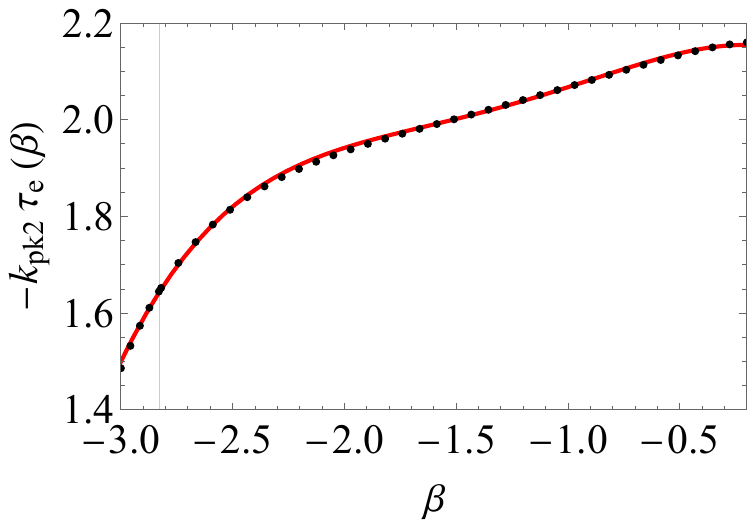}
\caption{Black dots: $-k_{\spk} \tau_\ue$ obtained from the truncated extremum condition, Eq.~\eqref{eq:5throot} without the last term. Red line: fitting formula in Eq.~\eqref{eq:fit2ndpk}. The vertical line indicates $\beta=-2.83$. The second-peak location is given by $\sqrt{x_1}$ for $\beta \gtrsim-2.83$ and by $\sqrt{x_2}$ for $\beta \lesssim-2.83$.}
\label{fig:2ndpkfit}
\end{figure}

We find that discarding the last term in Eq.~\eqref{eq:2ndpeakpl}, which we denote by $\mathcal{P}_{\mathcal{R}}^{\spk\text{-roots}}(k)$, still reproduces the second-peak position as a local maximum, as shown in Fig.~\ref{fig:solve-2nd-n2}.
After this truncation, the extremum condition reduces from the fifth-order equation in Eq.~\eqref{eq:5throot} to a quartic equation. 
In what follows, Eq.~\eqref{eq:2ndpeakpl} is used to reconstruct the spectrum around the second peak, while $\mathcal{P}_{\mathcal{R}}^{\spk\text{-roots}}(k)$ is used only to obtain analytic estimates of the second-peak location and amplitude.
Solving the quartic equation gives four roots. The two roots relevant for the second-peak location are
\begin{equation}\label{eq:x1x2}
x_1=r_1+\frac{1}{2} \sqrt{r_2}-\frac{1}{2} \sqrt{r_3 +r_4} , \quad 
x_2=r_1-\frac{1}{2} \sqrt{r_2}-\frac{1}{2} \sqrt{r_3 -r_4} .
\end{equation}
The auxiliary quantities $r_i$ ($i=1,2,3,4$) and the remaining two roots are listed in Eqs.~\eqref{eq:r1234} and \eqref{eq:x3x4}, respectively.
For $\beta \gtrsim -2.83$, the second peak is described by $-k_{\spk} \tau_\ue=\sqrt{x_1}$, while for $\beta \lesssim -2.83$, it is described by $-k_{\spk} \tau_\ue=\sqrt{x_2}$.
A fitting formula for $-k_{\spk} \tau_\ue$ is
\begin{equation}\label{eq:fit2ndpk}
-k_{\spk} \tau_\ue =2.14-0.145 \beta-0.406 \beta^2-0.235 \beta^3-0.0465 \beta^4\coloneqq  x_{\spk}(\beta),
\end{equation}
which is shown by the red solid line in Fig.~\ref{fig:2ndpkfit}.

To summarize, Eqs.~\eqref{eq:1stpkfit-nonatt}, \eqref{eq:1stpkfit-att}, and \eqref{eq:fit2ndpk} show that the positions of the two peaks are determined by the transition times.
According to Eq.~\eqref{app:14attra}, the amplitude of the first peak has no apparent dependence on the duration of the \ac{CR} stage in the attractor case. 
In the non-attractor case, however, the first-peak amplitude depends on $N_{\mathrm{CR}}$ through the overall factor $(\tau_\ue/\tau_\us)^{6+4\beta}$ in Eq.~\eqref{app:14}. 
The second-peak amplitude in Eq.~\eqref{eq:2ndpeakpl} always depends on $N_{\mathrm{CR}}$ through $A_\UV$.
The distinct $N_{\mathrm{CR}}$ dependence of the two peaks originates from the different superhorizon evolution of curvature perturbations during the \ac{CR} phase. 
In the attractor regime, the second mode decays outside the horizon, and the curvature perturbation rapidly approaches a constant after horizon exit. 
As a result, the amplitude of the first peak exhibits no strong sensitivity to the duration of the \ac{CR} stage. 
In contrast, in the non-attractor regime, the superhorizon curvature perturbation grows during the \ac{CR} phase, leading to an overall enhancement of the first-peak amplitude that scales with the duration $N_{\mathrm{CR}}$, encoded by the factor $(\tau_\ue/\tau_\us)^{6+4\beta}$ and consistent with the superhorizon behavior $\mathcal{R}_k\propto a^{-(3+2\beta)}$.
The second peak is associated with modes crossing the horizon near the end of the \ac{CR} phase, where the matching to the second \ac{SR} phase plays a crucial role. 
Consequently, its amplitude inherits a dependence on the duration of the \ac{CR} stage through the factor $(e^{-N_{\mathrm{CR}}})^{2\beta}$, regardless of whether the background is attractor or non-attractor.

\section{Applications}

In this section, we build on the results obtained in Sec.~\ref{s:Features} and discuss several applications. 
Suppose that two distinct peaks are observed, or constrained, in the curvature power spectrum. 
One may then ask whether they can be generated by an \ac{SR}--\ac{CR}--\ac{SR} scenario. 
We argue that this question can be addressed using the ratio of the two peak scales and the ratio of their amplitudes.
The corresponding model parameters, $\beta$ and $N_{\mathrm{CR}}$, can then be reconstructed using the method described in Sec.~\ref{s:solvebN}. 
If no consistent solutions exist, the observed two-peak structure cannot be explained within the parameter regime of the \ac{SR}--\ac{CR}--\ac{SR} model considered here.

Once the model parameters are reconstructed, the smoothed power spectrum derived in Sec.~\ref{s:IGW} can be used to estimate the associated \ac{SIGW} spectrum efficiently. 
These procedures are implemented in our open-source code~\cite{CRsolver}.

\subsection{Model Parameters}\label{s:solvebN}

\begin{figure}
\centering
\includegraphics[width=7.6cm]{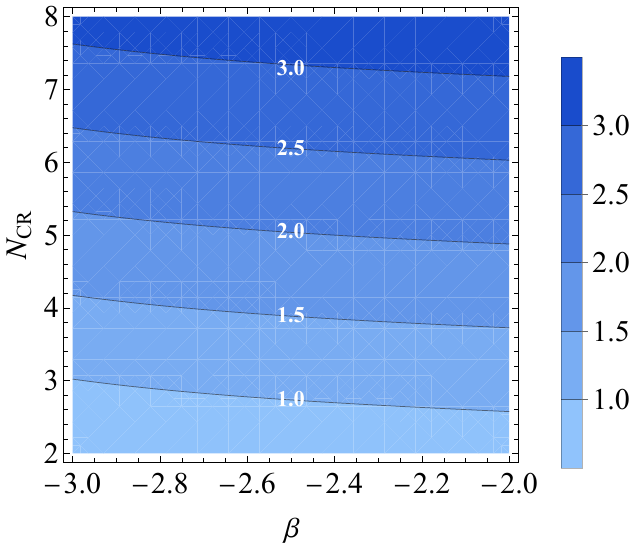}\includegraphics[width=7.3cm]{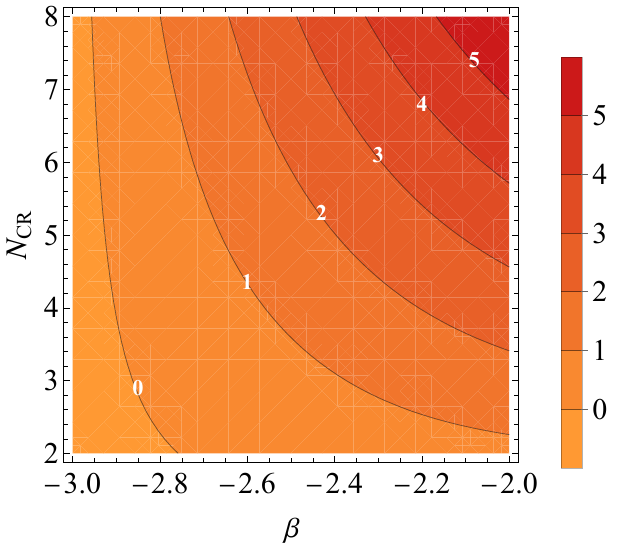}
\caption{Left: ratio of the second peak scale to the first peak scale, $\log_{10}(k_{\spk}/k_{\fpk})$, given by Eq.~\eqref{eq:k2ok1}. Right: ratio of the second peak amplitude to the first peak amplitude, $\log_{10}(A_{\spk}/A_{\fpk})$, given by Eq.~\eqref{eq:solvebeta}.}  
\label{fig:contour-ratkratP}
\end{figure}

Given the peak locations and amplitudes, we now consider how to infer $\beta$ and $N_{\mathrm{CR}}$ from $k_{\spk}/k_{\fpk}$ and $\mathcal{P}_{\mathcal{R}}^{\spk}/\mathcal{P}_{\mathcal{R}}^{\fpk}$ when the two peaks are both sufficiently high. 
Such a case occurs only for sufficiently negative values of $\beta$; otherwise, the first peak is significantly suppressed or even obscured by large oscillations, as shown in Fig.~\ref{fig:PRnumana}. 
The frequency ratio between the two peaks is given by Eq.~\eqref{eq:1stpkfit-nonatt} and Eq.~\eqref{eq:fit2ndpk},
\begin{equation}
\begin{aligned}\label{eq:k2ok1}
\frac{k_{\spk}}{k_{\fpk}}=e^{N_{\mathrm{CR}}}\frac{x_{\spk}(\beta)}{x_{\fpk}(\beta)}.
\end{aligned}
\end{equation}
For amplitude, we can estimate the amplitude of the second peak by substituting Eq.~\eqref{eq:fit2ndpk} into Eq.~\eqref{eq:2ndpeakpl},
\begin{equation}
\label{eq:amp2ndpk}
A_{\spk}
\approx
A_{\mathrm{IR}}\left(e^{-N_{\mathrm{CR}}}\right)^{2 \beta}~ f_2(\beta), \quad 
f_2 (\beta)\coloneqq  \sum_{n=0}^{6} d_{2n}(\beta) (-k_{\spk} \tau_\ue)^{2n}.
\end{equation}
Similarly, we substitute Eq.~\eqref{eq:1stpkfit-nonatt} into Eq.~\eqref{app:14} for the first peak, 
\begin{equation}
\label{eq:amp1stpk}
A_{\fpk}
\approx A_\IR\left(e^{-N_{\mathrm{CR}}}\right)^{6+4 \beta} f_1(\beta), \quad 
f_1(\beta) \coloneqq  \mathcal{W}_d^{-1}(\beta) (-k_{\fpk} \tau_\us)^4 \sum_{n=0}^7 \mathcal{W}_{2 n}(\beta) \left(-k_{\fpk} \tau_\us\right)^{2 n}.
\end{equation}

For phenomenological applications, we restrict the analysis in this subsection to $-3\leq\beta\lesssim -2$. 
In this regime, both peaks can be sufficiently pronounced and well separated, and the second peak is often comparable to or larger than the first. 
The relative hierarchy of the two peaks is important because it affects the characteristic scales of \ac{PBH} production and the shape of the associated induced gravitational-wave spectrum.

Under this assumption, we provide fitting formulae for $f_1(\beta)$ and $f_2(\beta)$:
\begin{equation}
\begin{aligned}\label{eq:fitf1f2}
f_1(\beta)&= 1637.551+2376.283 \beta+1310.923 \beta^2+323.051 \beta^3+29.949 \beta^4,\\
f_2(\beta)&= 51.354+89.605 \beta+60.791 \beta^2+18.264 \beta^3+2.086 \beta^4.
\end{aligned}
\end{equation}
Combining Eq.~\eqref{eq:amp2ndpk} and Eq.~\eqref{eq:amp1stpk}, one obtains
\begin{equation}
\begin{aligned}\label{eq:P2oP1}
\frac{A_{\spk}}{A_{\fpk}}=e^{(2\beta+6)N_{\mathrm{CR}}}\frac{f_2(\beta)}{f_1(\beta)}.
\end{aligned}
\end{equation}
Substituting Eq.~\eqref{eq:k2ok1} into Eq.~\eqref{eq:P2oP1}, we obtain 
\begin{equation}
\label{eq:solvebeta}
\frac{A_{\spk}}{A_{\fpk}}=\left(\frac{k_{\spk}}{k_{\fpk}} \frac{x_{\fpk}(\beta)}{x_{\spk}(\beta)}\right)^{2\beta+6} \frac{f_2(\beta)}{f_1(\beta)}.
\end{equation}

\begin{figure}
\centering
\includegraphics[width=7.9cm]{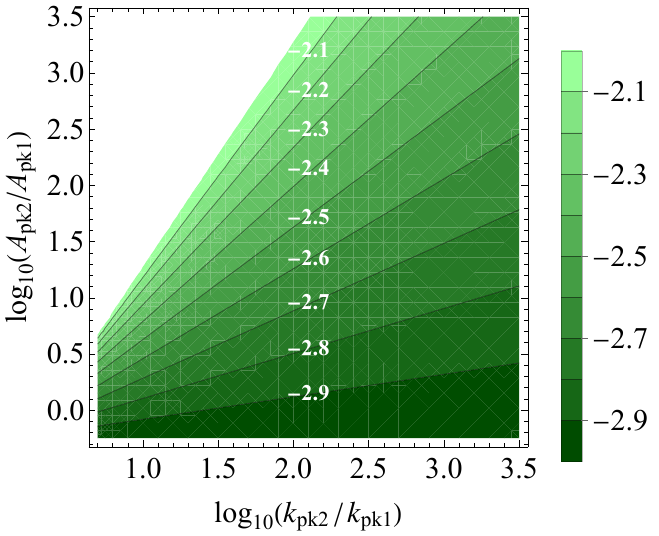}\includegraphics[width=7.3cm]{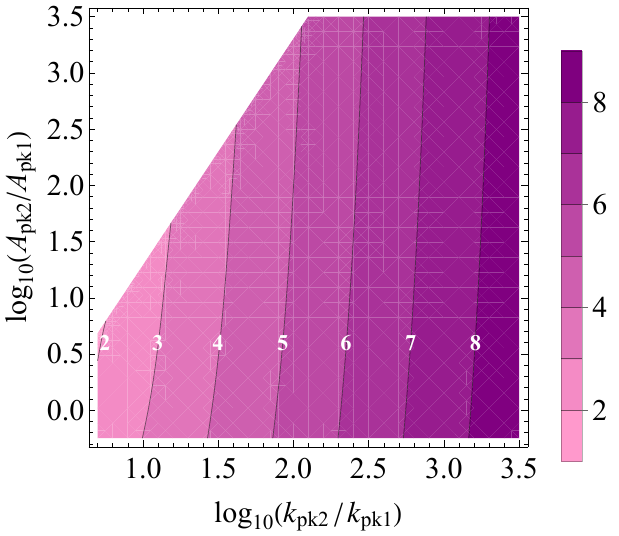}
\caption{Contour plot for $\beta$ (left) and $N_{\mathrm{CR}}$ (right) as a function of $\left( k_{\spk} / k_{\fpk}, A_{\spk} / A_{\fpk}\right)$, see Eq.~\eqref{eq:solvebeta}. The boundary line for $\beta=-2$ in the left panel is approximated by $\log_{10}(A_{\spk} / A_{\fpk})=-0.727 + 2.015\log_{10}(k_{\spk} / k_{\fpk})$.} 
\label{fig:contour-betaNCR}
\end{figure}

One can numerically determine $\beta$ from the ratios $\left( k_{\spk} / k_{\fpk}, A_{\spk} / A_{\fpk}\right)$ using Eq.~\eqref{eq:solvebeta}.
The value of $N_{\mathrm{CR}}$ can then be obtained from Eq.~\eqref{eq:k2ok1}. 
The resulting parameter reconstruction is illustrated in Fig.~\ref{fig:contour-betaNCR}. 
The boundary line can be understood as follows. 
The intermediate region between the first and second peaks is approximately described by $k^{6+2\beta}$, and hence the contour lines in the left panel have an approximate tilt of $6+2\beta$. 
For the boundary case $\beta=-2$, this slope becomes approximately $k^2$, consistent with the fitted boundary line shown in Fig.~\ref{fig:contour-betaNCR}.

After obtaining $\beta$ and $N_{\mathrm{CR}}$, one can determine $A_\IR$ using either Eq.~\eqref{eq:amp2ndpk} or Eq.~\eqref{eq:amp1stpk}. 
In practice, we use the larger peak amplitude to fix $A_\IR$ for the reconstruction.

\subsection{Implications for SIGW Spectrum}\label{s:IGW}

We now turn to the stochastic gravitational-wave background induced by scalar perturbations through the scalar--scalar--tensor interaction.
The induced GW spectrum we observe today is \cite{Pi:2020otn}
\begin{equation}
    \Omega_{\mathrm{GW}}(k) h^2=1.6 \times  10^{-5}\left(\frac{g_{*}\left(\tau_k\right)}{106.75}\right)^{-1 / 3}\left(\frac{\Omega_{\ur,0} h^2}{4.2 \times   10^{-5}}\right) \Omega_{\mathrm{GW}, \mathrm{eq}}(k),
    \label{eq: OGWh2}
\end{equation}
where $g_*(\tau_k)$ denotes the effective number of relativistic degrees of freedom of the radiation fluid at the horizon reentry of mode $k$. 
We approximate the effective degrees of freedom for energy and entropy density as equal. 
The quantity $\Omega_{\ur,0}h^2$ is the present radiation density, with the normalized Hubble constant $h=H_0/(\SI{100}{km/s/Mpc})$. 
The spectrum at matter-radiation equality, $\Omega_{\mathrm{GW}, \mathrm{eq}}(k)$, is given by~\cite{Kohri:2018awv}
\begin{equation}
    \begin{aligned}\label{eq:igw-general}
    \Omega_{\mathrm{GW}, \mathrm{eq}}(k) & =\frac{1}{12} \int_0^{\infty} \dd{t} \int_{-1}^1 \dd{s}\left[\frac{t(2+t)\left(s^2-1\right)}{(1-s+t)(1+s+t)}\right]^2 \mathcal{T}(s, t) \mathcal{P}_{\mathcal{R}}\qty(u k) \mathcal{P}_{\mathcal{R}}\qty(v k ), \\
    \mathcal{T}(s, t) 
    & \bmte{= \frac{288\left(-5+s^2+t(2+t)\right)^2}{(1-s+t)^6(1+s+t)^6}\left(\frac{\pi^2}{4}\left(-5+s^2+t(2+t)\right)^2 \Theta(t-(\sqrt{3}-1))\right. \\
     \left.+\left(-(t-s+1)(t+s+1)+\frac{1}{2}\left(-5+s^2+t(2+t)\right) \ln \left|\frac{-2+t(2+t)}{3-s^2}\right|\right)^2\right),}
    \end{aligned}
\end{equation}
where $u=(t+s+1) / 2$ and $v=(t-s+1) / 2$. 
Using the exact power spectrum in Eq.~\eqref{eq:oriPowSpe} directly includes many oscillatory details that are likely to be averaged out in realistic observations.
These details also slow down the calculation in \textit{Mathematica}. 
To simplify future analyses, we therefore propose a smoothed power spectrum constructed from Eq.~\eqref{app:14}, Eq.~\eqref{eq:2ndpeakpl}, and the UV limit in Eq.~\eqref{def:AUV}.

We now determine the matching scales for these three parts: the first-peak region, the second-peak region, and the UV limit.
Considering that the oscillation period in Eq.~\eqref{eq:2ndpeaktrig} is $\pi$~\cite{Pi:2022zxs}, the phase gap between a minimum and a maximum is approximately $\pi/2$, so we choose the cutting scale between the second-peak region and the UV limit as 
\begin{equation}\label{eq:cut2}
k_{\mathrm{cut,2}}\coloneqq  k_{\spk}-\frac
{\pi}{2\tau_\ue}.
\end{equation}
Similarly, we define the low-frequency cut-off
\begin{equation}\label{eq:cut1}
k_{\mathrm{cut,1}}\coloneqq k_{\fpk}-\frac
{\pi}{2\tau_\us},
\end{equation}
where $k_{\fpk}$ and $k_{\spk}$ are given by Eq.~\eqref{eq:1stpkfit-nonatt} and Eq.~\eqref{eq:fit2ndpk}, respectively.

\begin{figure}
\centering
\includegraphics[width=10cm]{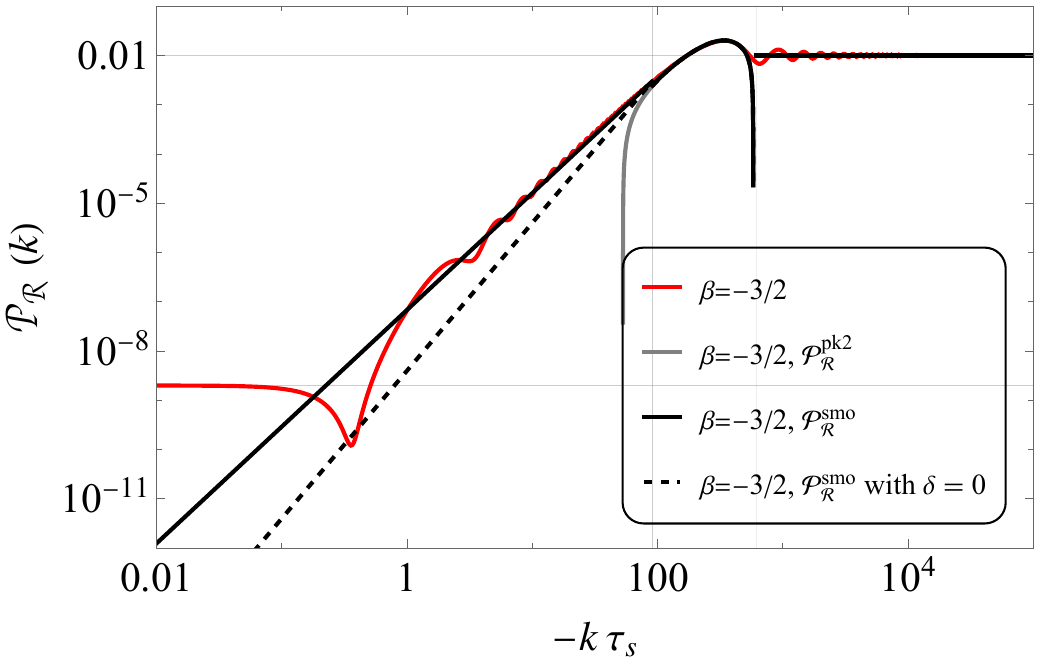}
\caption{Modification of the large-scale slope for $\beta\sim-3/2$. The vertical lines indicate $k_{\mathrm{mid}}$ and $k_{\mathrm{cut,2}}$. The smoothed spectrum $\mathcal{P}_{\mathcal{R}}^{\mathrm{smo}}$ is given by Eq.~\eqref{eq:Pmimic0}.}
\label{fig:b32-mimic}
\end{figure}

However, since Eq.~\eqref{eq:2ndpeakpl} was obtained from the UV expansion, it may become unreliable before reaching $k_{\mathrm{cut,1}}$ for some values of $\beta$. 
This can be handled by adding an intermediate power-law stage proportional to $k^{3-2\nu}$.
The first two terms in Eq.~\eqref{eq:2ndpeakpl} are sufficient to fix the matching point,
\begin{equation}
\mathcal{P}_{\mathcal{R}}^{\mathrm{mid}}(k) \coloneqq  A_\UV \sum_{n=0}^1 d_{2 n}(\beta)  ~\kappa_\ue^{2 n}.
\end{equation}
We then define the matching scale $k_{\mathrm{mid}}$ between the $k^{3-2\nu}$ power law and Eq.~\eqref{eq:2ndpeakpl} by
\begin{equation}
    \eval{\dv{\ln \mathcal{P}_{\mathcal{R}}^{\mathrm{mid}}(k)}{\ln k}}_{k_{\mathrm{mid}}}=3-2 \nu,
\end{equation}
which gives
\begin{equation}\label{eq:mid}
k_{\mathrm{mid}}=-\left(\frac{3-2\nu}{2\nu-1}\frac{d_0}{d_2}\right)^{1/2}\frac{1}{\tau_\ue}.
\end{equation}
Furthermore, as shown in Fig.~\ref{fig:b32-mimic}, for $\beta \approx -3/2$ one has to introduce an additional correction $k^\delta$ to the scaling $\mathcal{P}_{\mathcal{R}}(k) \propto k^{3-2\nu}$.
Numerically, if $A_\UV$ is fixed around $10^{-2}$, the power modulation $\delta(\beta)$ is well fitted by 
\begin{equation}\label{def:delta}
\delta(\beta)\approx
\begin{cases}
-0.63-1.44\left(\beta+\dfrac{3}{2}\right)-0.43\left(\beta+\dfrac{3}{2}\right)^2, & -2<\beta \le -3/2, \\
-0.63+2.50\left(\beta+\dfrac{3}{2}\right)-2.90\left(\beta+\dfrac{3}{2}\right)^2,  & -3/2 <\beta <-1, \\
0 & \text{otherwise}.
\end{cases}
\end{equation}
For modes crossing the horizon during the \ac{CR} phase with $\beta$ values near the attractor/non-attractor boundary $\beta=-3/2$, the superhorizon evolution is extremely slow. 
As a result, the curvature power spectrum $\mathcal{P}_{\mathcal{R}}(k)$ acquires small corrections and deviates from the idealized $k^{3-2\nu}$ power law. 
We leave the detailed study of this boundary effect to future work. For the cases with $-3\leq\beta\lesssim-2$ considered here, the power modulation is expected to be small.

\begin{figure}
    \centering
    \includegraphics[width=7.6cm]{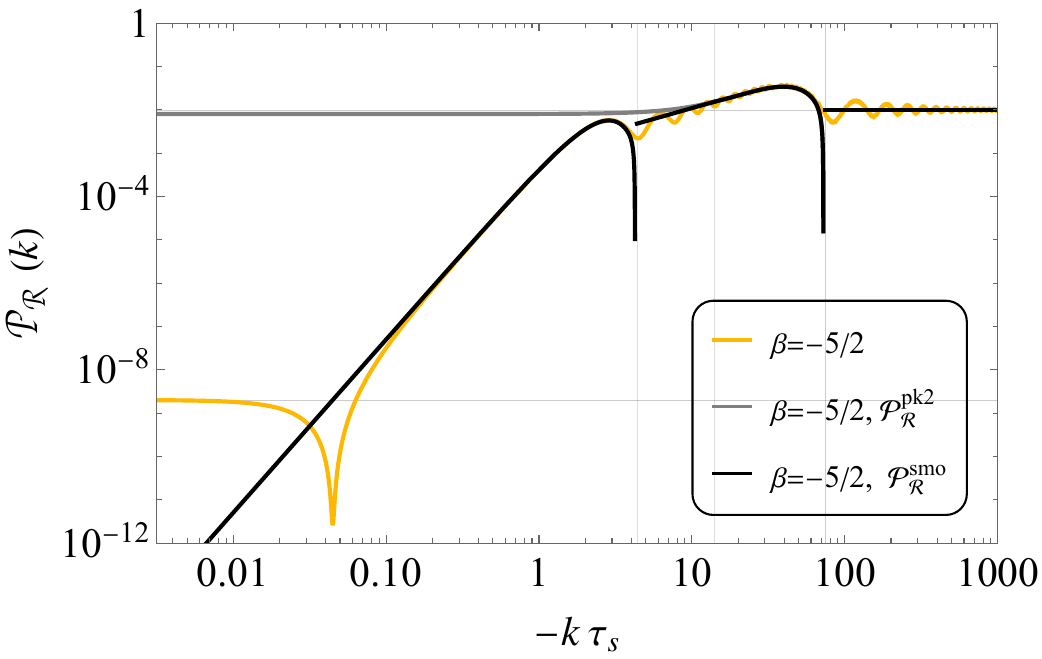}
    \includegraphics[width=7.6cm]{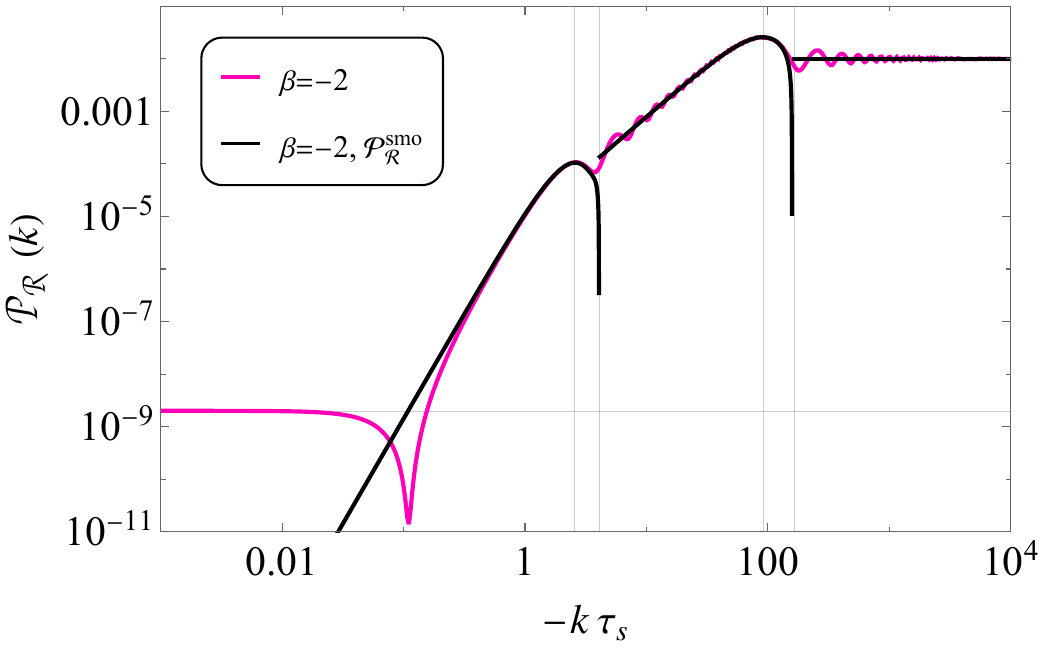}
    \caption{Left: Eq.~\eqref{eq:2ndpeakpl} (gray) becomes unreliable on the large-scale side for $\beta=-5/2$. The vertical lines indicate $k_{\mathrm{cut,1}}$, $k_{\mathrm{mid}}$, and $k_{\mathrm{cut,2}}$, from left to right. Right: smoothed power spectrum for $\beta=-2$, given by Eq.~\eqref{eq:Pmimic}. The vertical lines indicate $k_{\fpk}$, $k_{\mathrm{cut,1}}$, $k_{\spk}$, and $k_{\mathrm{cut,2}}$, from left to right.}
    \label{fig: approximate power}
\end{figure}

\begin{figure}
    \centering
    \includegraphics[width=12cm]{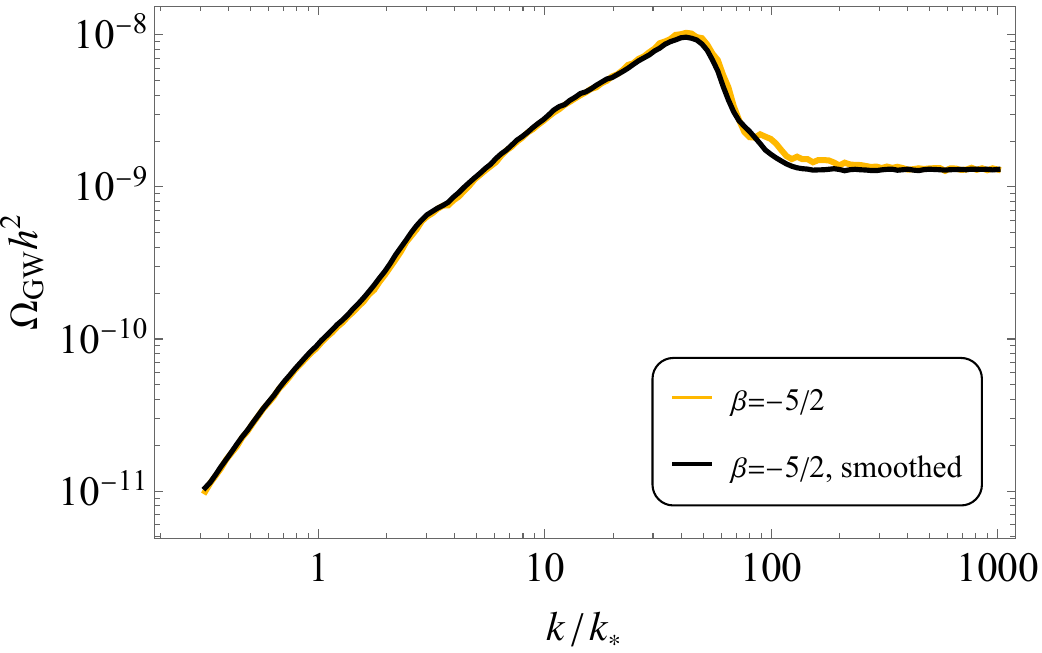}
    \caption{\ac{SIGW} signals for $\beta=-5/2$ 
    and $A_\UV=10^{-2}$. 
    The yellow line corresponds to
    the original power spectrum, while the black line is obtained
    from the smoothed power spectrum, Eq.~\eqref{eq:Pmimic}.
    $k_* \coloneqq  -1/\tau_\us$ is the comoving wavenumber that crossed the horizon at $\tau_\us$.}
    \label{fig: OGW}
\end{figure}

Therefore, we propose the following final smoothed power spectrum for $-3 \leq \beta \lesssim-2$:
\bae{
        \label{eq:Pmimic0}\calP_\calR^{\mathrm{smo}}(k) \coloneqq \bde{
            \calP_\calR^\fpk(k), & k<k_{\cut,1}, \\
            A_\umid\pqty{\frac{k}{k_\umid}}^{3-2\nu }, & k_{\cut,1}\leq k<k_\umid, \\
            \calP_\calR^\spk(k), & k_\umid\leq k<k_{\cut,2}, \\
            A_\UV, & k_{\cut,2}\leq k,
        }
    }
where
\begin{equation}
    A_\umid \coloneqq \mathcal{P}_{\mathcal{R}}^{\mathrm{mid}}(k_\umid),
\end{equation}
if $k_\umid > k_{\cut,1}$. For $k_\umid\leq k_{\cut,1}$ (see the right panel of Fig.~\ref{fig: approximate power}),
    \bae{
        \label{eq:Pmimic}
        \calP_\calR^{\mathrm{smo}}(k) \coloneqq \bde{
            \calP_\calR^\fpk(k), & k<k_{\cut,1}, \\
            \calP_\calR^\spk(k), & k_{\cut,1}\leq k<k_{\cut,2}, \\
            A_\UV, & k_{\cut,2}\leq k.
        }
    }
For $\beta \approx -3/2$, the contribution from Eq.~\eqref{def:delta} should also be taken into account in Eq.~\eqref{eq:Pmimic0}, as shown in Fig.~\ref{fig:b32-mimic}.
Figure~\ref{fig: OGW} compares the \ac{SIGW} spectra obtained from the exact and smoothed curvature power spectra. 

\begin{figure}
\centering
\includegraphics[width=12cm]{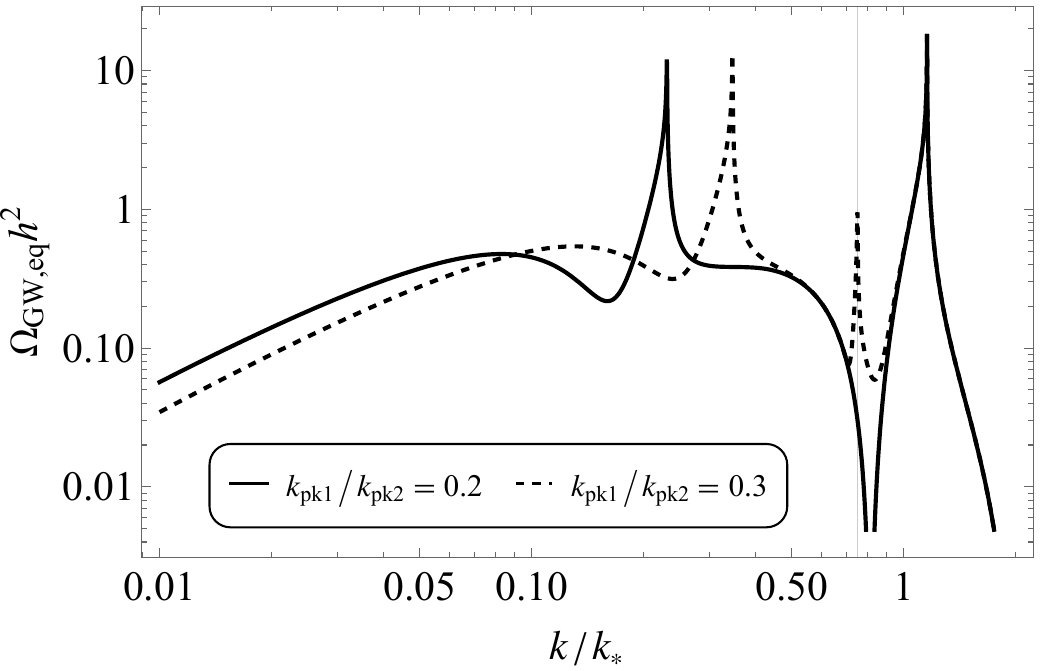}
\caption{\ac{SIGW} generated by Eq.~\eqref{eq:dpmpno}. $k_{\spk}/k_*=1$ 
and $A_{\fpk}=A_{\spk}=1$. 
No resonance is indicated for $k_{\fpk} / k_{\spk}=0.2$.
If $k_{\fpk}/k_{\spk} \lesssim 0.3$, the resonance peak disappears. The resonance peak for $k_{\fpk} / k_{\spk}=0.3$ is located at $k/k_* \approx 0.75$, as shown by the vertical line.}
\label{fig:0d3}
\end{figure}

\begin{figure}
\centering
\includegraphics[width=7.8cm]{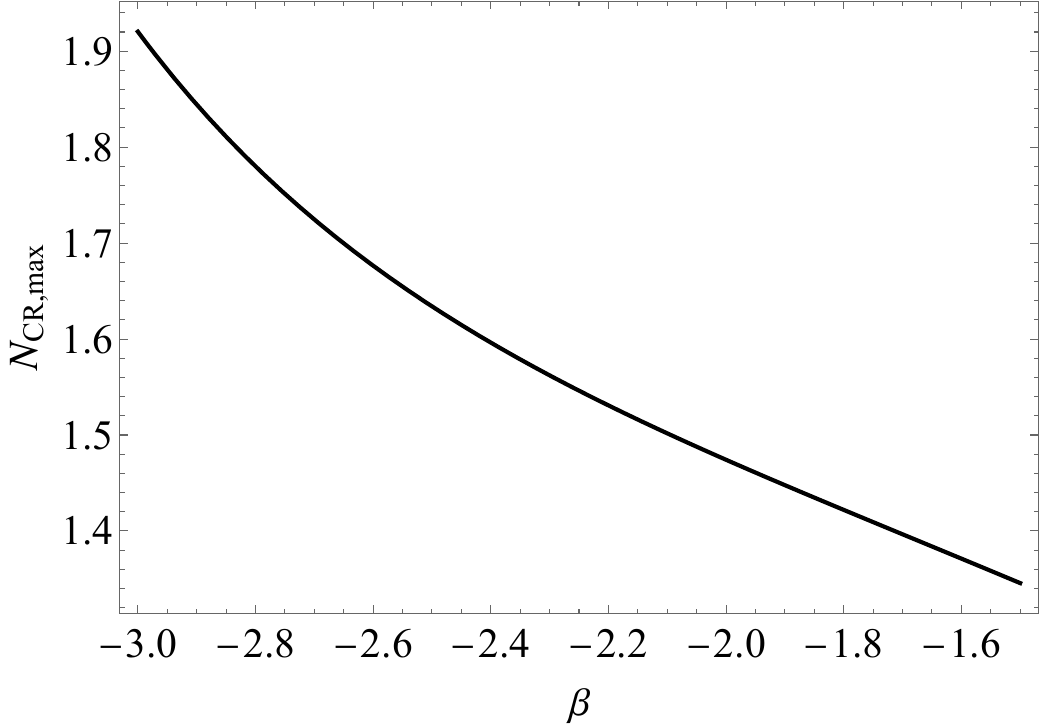}\includegraphics[width=7.8cm]{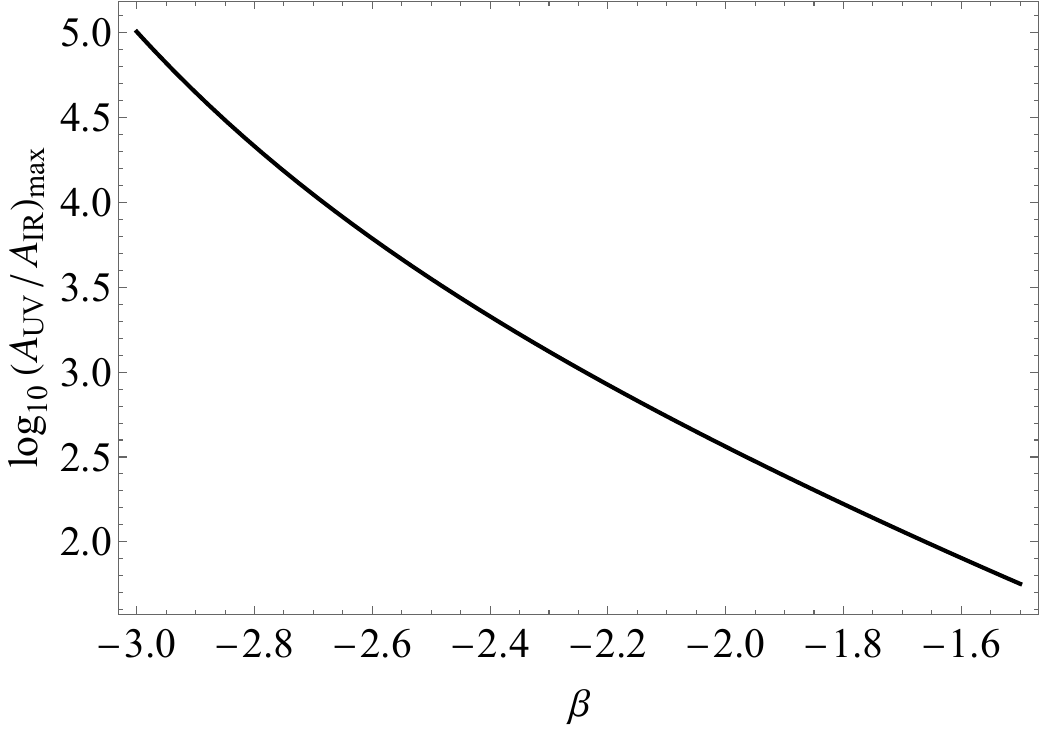}
\caption{Maximum values of $N_{\mathrm{CR}}$ and $A_\UV / A_\IR$ obtained by imposing $k_{\fpk} = 0.3 k_{\spk}$; see Eqs.~\eqref{eq:NCRmax} and \eqref{eq:maxUV}.}
\label{fig:maxenh}
\end{figure}

\begin{figure}
\centering
\includegraphics[width=7.6cm]{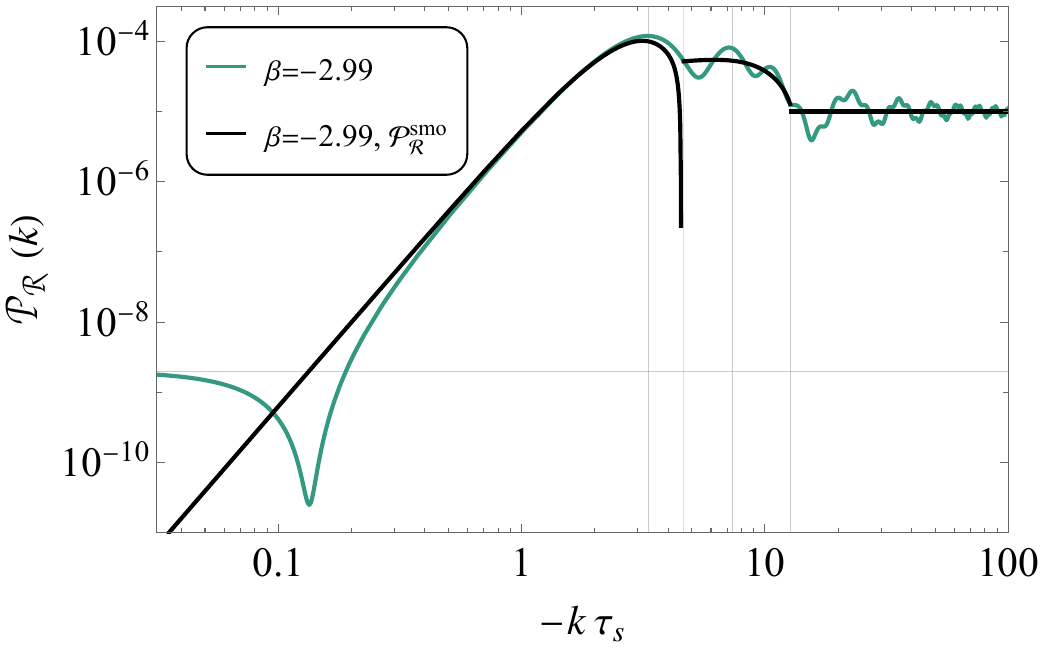}\includegraphics[width=8.1cm]{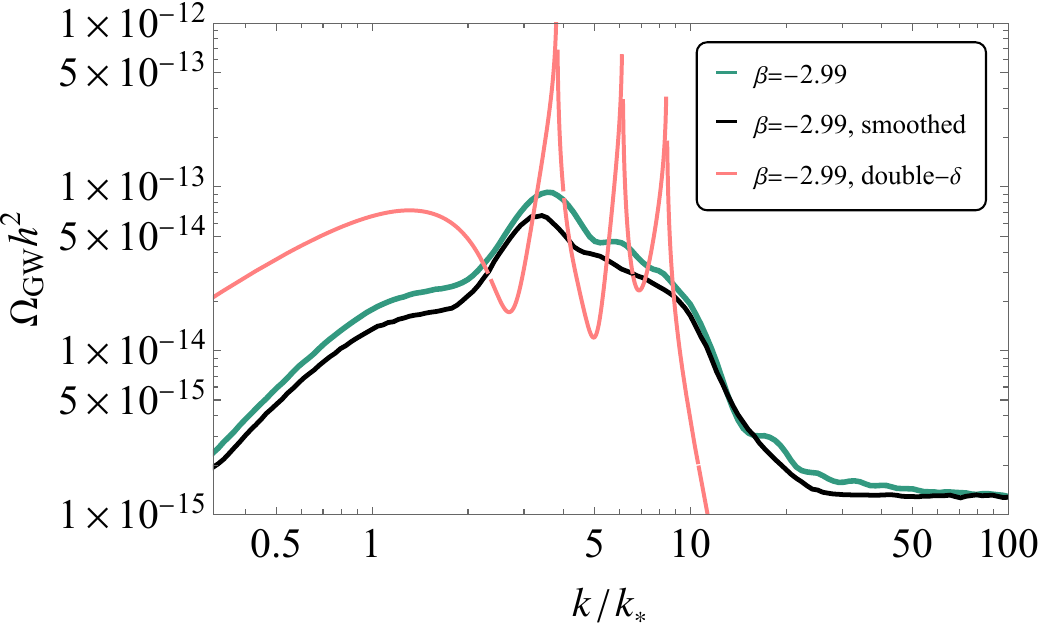}
\caption{Left: power spectrum for $\beta=-2.99$, $A_\UV=10^{-5}$. The vertical lines are $\kappa_\us =3.3$, $\kappa_\us =k_{\mathrm{cut,1}}$, $\kappa_\us= 7.3$, and $\kappa_\us =k_{\mathrm{cut,2}}$. Here,
$\kappa_\us =3.3$ and $\kappa_\us =7.3$ are the two scales with the largest amplitudes.
Right: \ac{SIGW} spectrum from exact power spectrum, smoothed power spectrum and double-$\delta$ power spectrum Eq.~\eqref{eq:dpmpno}. 
The smoothed power spectrum is lower than the exact one, since our approximate formulae are more accurate when $A_\UV/A_\IR$ is large.
One can also find an additional peak (the one in the middle) predicted at the same position, arising from the cross term in the double-$\delta$ power spectrum approximation.}
\label{fig:exm29}
\end{figure}

Another potentially interesting aspect is the resonance peak in the \ac{SIGW} spectrum that appears as the two peaks approach each other. 
This phenomenon was previously explored with a doubly monochromatic power spectrum~\cite{Cai:2019amo}. 
Substituting the two peaks from our model,
\begin{equation}\label{eq:dpmpno}
\mathcal{P}_{\mathcal{R}}^{\text{double-}\delta}(k) \coloneqq  A_{\fpk}\delta_{\mathrm{D}}(\ln\tilde{k}_1)+A_{\spk}\delta_{\mathrm{D}}(\ln\tilde{k}_2),
\end{equation}
where $\tilde{k}_1 \coloneqq k / k_{\fpk}$ and $\tilde{k}_2 \coloneqq k / k_{\spk}$. 
Using the formula in Ref.~\cite{Cai:2019amo}, we find that the resonance structure appears when $0.3~ k_\spk \lesssim k_\fpk \lesssim k_\spk$ for $A_\fpk = A_\spk$; see Fig.~\ref{fig:0d3}. 
For $A_\fpk \ll A_\spk$, the second peak is expected to dominate over any potential resonance signal. 
Combining Eq.~\eqref{eq:1stpkfit-nonatt} and Eq.~\eqref{eq:fit2ndpk}, and imposing $k_{\fpk} = 0.3 k_{\spk}$, the maximum allowed value of $N_{\mathrm{CR}}$ is
\begin{equation}\label{eq:NCRmax}
N_{\mathrm{CR,max}}(\beta)=-\ln (\frac{0.3 x_{\spk} (\beta)}{x_{\fpk} (\beta)}).
\end{equation} 
The maximum enhancement in the UV limit is  
\begin{equation}\label{eq:maxUV}
\left.\frac{A_\UV}{A_\IR}\right|_{\mathrm{max}}(\beta)=\exp(-2\beta N_{\mathrm{CR,max}}(\beta)).
\end{equation} 
We show Eq.~\eqref{eq:NCRmax} and Eq.~\eqref{eq:maxUV} in Fig.~\ref{fig:maxenh}.
The maximum enhancement is $\sim 10^{5}$, and therefore $A_\UV\sim 10^{-4}$ if one adopts the COBE normalisation for the IR amplitude as $A_\IR\sim2\times10^{-9}$.
This enhancement is not sufficient to generate a large-amplitude \ac{SIGW} signal; the resulting amplitude would be around $10^{-13}$.
Moreover, when $N_{\mathrm{CR}}$ is quite small, Eq.~\eqref{eq:1stpkfit-nonatt} and Eq.~\eqref{eq:fit2ndpk} may no longer provide a proper description of the spectrum, and the crests of the oscillations between the two peaks can be more important than the envelope around the ``second peak''.
By \textit{crest}, we refer to the local maxima points arising from mode mixing, whereas \textit{peak} refers to global maxima generated by the two transitions at $\tau=\tau_\us$ and $\tau=\tau_\ue$. 
For example, for $\beta=-2.99$ and $A_\UV=10^{-5}$, there is a mild resonance peak between the first peak and the highest nearby crest, as shown in Fig.~\ref{fig:exm29}. 
The highest crest around the first peak actually comes from the oscillation generated by the first transition and is not directly related to the second transition. 
Therefore, we conclude that the resonance peak in the \ac{SIGW} spectrum produced by the two transition-induced peaks at $\tau_\us$ and $\tau_\ue$ may be difficult to observe. 

\section{Conclusion and Discussion}\label{s:conclusion}

Constant-roll inflation provides a theoretically controlled extension of standard slow-roll inflation, for which many analytic results are available and phenomenological applications have been explored.
In this work, we have carried out a detailed analysis of the transient \ac{SR}--\ac{CR}--\ac{SR} model, a minimal yet flexible framework for describing temporary departures from slow-roll evolution. 
By focusing on a transient \ac{CR} phase bounded by two sharp transitions, we have identified and characterized the distinctive features of the curvature power spectrum, in particular the positions and amplitudes of the two peaks associated with the two transitions.

A central result of this work is that, in the regime where the two peaks are sufficiently prominent and well separated, these spectral features contain enough information to reconstruct the parameters of the \ac{CR} phase without a brute-force scan. 
This reconstruction is applicable in the regime $-3 \leq \beta \lesssim -2$, where the spectrum develops two well-separated global peaks, with the second peak comparable to or typically larger than the first.
The resulting inversion procedure provides a practical way to interpret future measurements or constraints on enhanced small-scale power spectra and to discriminate \ac{CR} dynamics from other mechanisms that generate similar enhancements.

We have also introduced an analytic formula for the curvature power spectrum that retains the main peak structure while averaging out rapidly oscillating features. 
This approximation substantially simplifies the estimation of the corresponding \ac{SIGW} spectrum and enables us to clarify when the resonance feature associated with two nearby scalar peaks can become relevant. 
In the parameter range considered here, however, this resonance effect is typically subdominant and therefore difficult to observe.

Beyond these immediate applications, our results show that \ac{CR} inflation can serve as a useful building block for modelling controlled, transient departures from slow roll in multi-phase inflationary scenarios. 
The analytic formulae and computational shortcuts developed here should facilitate more efficient phenomenological studies of \ac{CR}-based scenarios and their induced gravitational-wave signatures.

\section*{Acknowledgements}
This work is supported by the National Key Research and Development Program of China Grant No.\ 2021YFC2203004 
and by JSPS KAKENHI Grant Nos.~JP22K03639 (H.M.), JP24K00624 (S.P.\ and J.W.), and JP24K07047 (Y.T.).
S.P.\ is also supported in part by the National Natural Science Foundation of China (NSFC) Grants No.\ 12475066 and No.\ 12447101.
J.W.\ is also supported by Kavli IPMU, which was established by the World Premier International Research Center Initiative (WPI), MEXT, Japan.

\appendix

\section{Coefficients in Main Text}\label{s:coeff}

Coefficients in Eq.~\eqref{app:14}:
\begin{equation}
\begin{aligned}
\mathcal{W}_d(\beta)&=540(8\beta^2)^{-1} (-9+2 \beta)(-7+2 \beta)(-5+2 \beta)\left(9-40 \beta^2+16 \beta^4\right)^2 ,\\
\mathcal{W}_0(\beta)&=120 \beta^2(-9+2 \beta)(-7+2 \beta)(-5+2 \beta)(3+4(-2+\beta) \beta)^2 ,\\
\mathcal{W}_2(\beta)&=30(-9+2 \beta)(-7+2 \beta)(-5+2 \beta)(3-2 \beta)^2(-1+2 \beta)\left(-1-2 \beta+4 \beta^2\right) ,\\
\mathcal{W}_4(\beta)&=60 \beta(-9+2 \beta)(-7+2 \beta)(-5+2 \beta)(-3+2 \beta)(1+2(-2+\beta) \beta) ,\\
\mathcal{W}_6(\beta)&=10 \beta(-9+2 \beta)(-7+2 \beta)(-3+2 \beta)(9+2 \beta(-7+2 \beta)) ,\\
\mathcal{W}_8(\beta)&=20(-2+\beta) \beta(-9+2 \beta)(5+(-5+\beta) \beta), \\
\mathcal{W}_{10}(\beta)&=(-3+\beta) \beta\left(35-26 \beta+4 \beta^2\right) ,\\
\mathcal{W}_{12}(\beta)&=\frac{4 \beta ^5-60 \beta ^4+326 \beta ^3-762 \beta ^2+648 \beta }{3 (2 \beta -11) (2 \beta -5)}, \\
\mathcal{W}_{14}(\beta)&=\frac{4 \beta ^5-74 \beta ^4+499 \beta ^3-1453 \beta ^2+1540 \beta }{21 (2 \beta -13) (2 \beta -11) (2 \beta -5)} .
\end{aligned}\label{eq:Yi}
\end{equation}

Coefficients in Eq.~\eqref{app:14attra}:
\begin{equation}
\begin{aligned}\label{eq:appattr-18}
\mathcal{Y}_0(\beta)&=\prod_{n=1}^{4}\left(2n+1+2 \beta \right)^2 \prod_{n=5}^{9}\left(2n+1+2 \beta \right) ,\\
\mathcal{Y}_2(\beta)&=2\beta (3+2\beta)^{-1}~\mathcal{Y}_0(\beta), \\
\mathcal{Y}_4(\beta)&=-\beta (5+2 \beta) (9+4 \beta)\prod_{n=3}^{4}\left(2n+1+2 \beta \right)^2 \prod_{n=5}^{9}\left(2n+1+2 \beta \right) ,\\
\mathcal{Y}_6(\beta)&=\frac{2}{3} \beta (9+2 \beta)(10+3 \beta) \prod_{n=2}^{9}\left(2n+1+2 \beta \right) ,\\
\mathcal{Y}_8(\beta)&=-\frac{\beta}{6} (3+\beta) (35+8 \beta)\prod_{n=3}^{9}\left(2n+1+2 \beta \right)  ,\\
\mathcal{Y}_{10}(\beta)&= \frac{\beta}{15} (4+\beta) (27+5 \beta) (11+2\beta)^{-1} \prod_{n=3}^{9}\left(2n+1+2 \beta \right), \\
\mathcal{Y}_{12}(\beta)&= -\frac{\beta}{90} (4+\beta)(5+\beta) (77+12 \beta)(9+2 \beta)\prod_{n=7}^{9}\left(2n+1+2 \beta \right) ,\\
\mathcal{Y}_{14}(\beta)&= \frac{\beta}{315} (5+\beta)(6+\beta) (9+2 \beta)(52+7 \beta)\prod_{n=8}^{9}\left(2n+1+2 \beta \right) .
\end{aligned}
\end{equation}

Coefficients in Eq.~\eqref{eq:2ndpeaktrig}:
\begin{equation}
\begin{aligned}
\mathcal{S}_\us &=\bmte{2^{-14}(\beta -1)^2 \beta ^2 (\beta +1)^2 (\beta +2)^2 \kappa_\us^{-6} +2^{-8} \beta ^2 \kappa_\us^{-2}\\
+2^{-12} \beta ^2 (\beta +1) \left(-16+21 \beta^2+19 \beta^3+7 \beta^4+\beta^5\right) \kappa_\us^{-4} +2^{-6},}\\
\mathcal{A}&=\bmte{\Biggl\{ 2^{-2} \beta ^2 \biggl[(\beta -1)^4 \beta ^2 (\beta +1)^4 (\beta +2)^4 \kappa_\ue^{-12} + 2^{14}\kappa_\ue^{-2}\\
+ 2^{12} \left(-1+3 \beta+9 \beta^2+5 \beta^3+\beta^4\right) \kappa_\ue^{-4}\\
+ 2^{8} (\beta +1) \left(4-27 \beta^2+15 \beta^3+40 \beta^4+24 \beta^5+7 \beta^6+\beta^7\right) \kappa_\ue^{-6}\\
+ 2^{6} \beta ^2 (\beta +1)^2 \left(24-72 \beta-91 \beta^2+53 \beta^3+155 \beta^4+124 \beta^5+51 \beta^6+11 \beta^7+\beta^8 \right) \kappa_\ue^{-8}\\
+ 2^{2} \beta ^2 (\beta +1)^3 \left(\beta ^2+\beta -2\right)^2 \left(-20+16 \beta+37 \beta^2+23 \beta^3+7 \beta^4+\beta^5 \right) \kappa_\ue^{-10}\biggr] \Biggr\}^{1/2},}\\
\Phi &=\arctan\qty[\frac{(\beta -1)^2 \beta  (\beta +1)^2 (\beta +2)^2-2^6 \kappa_\ue^{4} +2^3 \beta  (\beta +1) (\beta +3) \left(-2+2 \beta+3 \beta^2+\beta^3\right) \kappa_\ue^{2}}{2^7 \kappa_\ue^{5} +2^4 (\beta +1) \left(-2+5 \beta+4 \beta^2+\beta^3\right) \kappa_\ue^{3}-2 (\beta -1)^2 \beta  (\beta +1)^2 (\beta +2)^2 \kappa_\ue}],\\
\theta &=\bmte{2^{-1}  (\beta -1)^2 \beta ^2 (\beta +1)^2 (\beta +2)^2 \kappa_\ue^{-6} +2^{6} +2^{5} \beta ^2 \kappa_\ue^{-2}\\
+ \beta ^2 (\beta +1) \left(-20+16 \beta+37 \beta^2+23 \beta^3+7 \beta^4+\beta^5 \right) \kappa_\ue^{-4}.}
\end{aligned}\label{eq:2ndpeaktrigcoeff}
\end{equation}

Coefficients in Eq.~\eqref{eq:2ndpeakpl}:
\begin{equation}
\begin{aligned}
d_0&= 1+\frac{4 \beta}{3}-\frac{19 \beta^2}{72}-\frac{103 \beta^3}{72}-\frac{413 \beta^4}{288}-\frac{61 \beta^5}{72}-\frac{43 \beta^6}{144}-\frac{\beta^7}{18}-\frac{\beta^8}{288},\\
d_2&= \frac{\beta}{2880}\left(-3072+196 \beta+836 \beta^2+713 \beta^3+360 \beta^4+114 \beta^5+20 \beta^6+\beta^7\right),\\
d_4&= \frac{\beta}{50400}\left(11520-364 \beta-1404 \beta^2-1093 \beta^3-492 \beta^4-142 \beta^5-24 \beta^6-\beta^7\right),\\
d_6&= \frac{\beta}{1360800}\left(-30720+580 \beta+2116 \beta^2+1553 \beta^3+640 \beta^4+170 \beta^5+28 \beta^6+\beta^7\right),\\
d_8&= \frac{\beta}{52390800} \left(67200-844 \beta-2972 \beta^2-2093 \beta^3-804 \beta^4-198 \beta^5-32 \beta^6-\beta^7\right),\\
d_{10}&= \frac{\beta}{2724321600} \left(-129024+1156 \beta+3972 \beta^2+2713 \beta^3+984 \beta^4+226 \beta^5+36 \beta^6+\beta^7\right),\\
d_{12}&= \frac{\beta}{183891708000} \left(225792-1516 \beta-5116 \beta^2-3413 \beta^3-1180 \beta^4-254 \beta^5-40 \beta^6-\beta^7\right).
\end{aligned}\label{eq:around2ndpkcoeff}
\end{equation}

$r_i$ ($i=1,2,3,4$) in Eq.~\eqref{eq:x1x2}:
\begin{equation}
\begin{aligned}\label{eq:r1234}
a&=3 \qty(20 d_{10}d_{2} +3d_{6}^2-8 d_{4}    d_{8} ) , \\
b&=54 \qty(10 d_{10}d_{4}^2-20 d_{10}   d_{2}    d_{6} +d_{6}^3-4 d_{4}    d_{6}    d_{8} +8 d_{2} d_{8}^2)  ,\\
r_1&= -\frac{d_8}{5 d_{10}} ,\\
r_2&=-\frac{2 d_{6} }{5 d_{10}}+\frac{4d_{8}^2}{25(d_{10})^2} +\frac{ 2^{1 / 3} a }{15  d_{10} \left(b +\sqrt{ -4 a^3+b^2}\right)^{1 / 3}  }+\frac{1}{15 \times  2^{1 / 3}  d_{10} }\left(b +\sqrt{ -4 a^3+b^2}\right)^{1 / 3} ,   \\
r_3&=-\frac{4 d_{6} }{5 d_{10}}+\frac{8d_{8}^2}{25(d_{10})^2} -\frac{ 2^{1 / 3} a  }{15  d_{10} \left(b +\sqrt{ -4 a^3+b^2}\right)^{1 / 3} }-\frac{1}{15 \times  2^{1 / 3}  d_{10} }\left(b +\sqrt{ -4 a^3+b^2}\right)^{1 / 3} ,   \\
r_4&=\left.\qty(-\frac{4 d_{4} }{5  d_{10} }+\frac{12 d_{6}    d_{8} }{25d_{10}^2}-\frac{16 d_{8}^3}{125 d_{10}^3}) \middle/\sqrt{r_2} \right..
\end{aligned}
\end{equation}

Other roots of Eq.~\eqref{eq:fit2ndpk} without the last term:
\begin{equation}\label{eq:x3x4}
x_3=r_1+\frac{1}{2} \sqrt{r_2 }+\frac{1}{2} \sqrt{r_3 +r_4} , \quad 
x_4=r_1-\frac{1}{2} \sqrt{r_2 }+\frac{1}{2} \sqrt{r_3 -r_4} .
\end{equation}

\bibliography{ref}
\addcontentsline{toc}{section}{Bibliography}
\bibliographystyle{JHEP}

\end{document}